\documentclass[usenatbib]{mn2e}

\usepackage{amsmath,amssymb, graphics}
\usepackage{pgf,tikz}
\usepackage{bm}
\usepackage{graphicx}
\usepackage{epstopdf}
\usepackage{float}
\usepackage{mathrsfs}
\usepackage{longtable}
\allowdisplaybreaks

\newcommand{\bT}{{\bm T}}
\newcommand{\bbT}{\tilde {\bm T}}

\newcommand{\boms}{\tilde \omega_{\rm ds}}
\newcommand{\bomb}{\tilde \omega_{\rm db}}
\newcommand{\bomd}{\tilde \omega_{\rm sd}}

\newcommand{\omps}{\omega_{\rm ps}}
\newcommand{\bompd}{\tilde \omega_{\rm pd}}
\newcommand{\bompdg}{\tilde \omega_{\rm pd>}}
\newcommand{\bompdl}{\tilde \omega_{\rm pd<}}
\newcommand{\ompb}{\omega_{\rm pb}}
\newcommand{\bomsd}{\tilde \omega_{\rm sd}}
\newcommand{\bomsdg}{\tilde \omega_{\rm sd>}}
\newcommand{\bomsdl}{\tilde \omega_{\rm sd<}}
\newcommand{\omsp}{\omega_{\rm sp}}
\newcommand{\bomds}{\tilde \omega_{\rm ds}}
\newcommand{\bomdgs}{\tilde \omega_{\rm d>s}}
\newcommand{\bomdls}{\tilde \omega_{\rm d<s}}
\newcommand{\bomdp}{\tilde \omega_{\rm dp}}
\newcommand{\bomdgp}{\tilde \omega_{\rm d>p}}
\newcommand{\bomdlp}{\tilde \omega_{\rm d<p}}
\newcommand{\bomdb}{\tilde \omega_{\rm db}}
\newcommand{\tp}{t_{\rm p}}

\newcommand{\ME}{\text{M}_\oplus}
\newcommand{\MJ}{\text{M}_{\rm J}}
\newcommand{\Myr}{\text{Myr}}

\newcommand{\bs}{\bm {\hat s}}
\newcommand{\bl}{\bm {\hat l}}

\newcommand{\rin}{r_\text{in}}
\newcommand{\rout}{r_\text{out}}
\newcommand{\Sgout}{\Sigma_\text{out}}
\newcommand{\Md}{M_\text{d}}
\newcommand{\Ld}{L_\text{d}}
\newcommand{\bLd}{{\bm L}_\text{d}}
\newcommand{\der}{\text{d}}

\newcommand{\Om}{\Omega}
\newcommand{\om}{\omega}
\newcommand{\Dg}{\Delta}

\newcommand{\Lam}{\Lambda}

\newcommand{\tg}{\theta}

\newcommand{\tgsd}{\theta_{\rm sd}}
\newcommand{\tgsb}{\theta_{\rm sb}}
\newcommand{\tgdb}{\theta_{\rm db}}

\newcommand{\blb}{\bm {\hat l}_\text{b}}

\newcommand{\bcdot}{{\bm \cdot}}
\newcommand{\btimes}{{\bm \times}}

\newcommand{\ag}{\alpha}

\newcommand{\Sg}{\Sigma}

\newcommand{\tw}{t_{\rm w}}
\newcommand{\tv}{t_{\rm v}}
\newcommand{\tvout}{t_{\rm v, out}}
\newcommand{\tvin}{t_{\rm v, in}}
\newcommand{\rc}{r_{\rm c}}
\newcommand{\Sgc}{\Sigma_{\rm c}}

\newcommand{\brin}{\bar r_\text{in}}
\newcommand{\brout}{\bar r_\text{out}}
\newcommand{\bMd}{\bar M_\text{d}}

\newcommand{\bS}{{\bm S}}
\newcommand{\Mb}{M_{\rm b}}
\newcommand{\bMb}{\bar M_{\rm b}}
\newcommand{\ab}{a_{\rm b}}
\newcommand{\bab}{\bar a_{\rm b}}
\newcommand{\Mst}{M_\star}
\newcommand{\Ms}{M_\star}
\newcommand{\bMst}{\bar M_\star}
\newcommand{\bMs}{\bar M_\star}
\newcommand{\Rst}{R_\star}
\newcommand{\Rs}{R_\star}
\newcommand{\bRst}{\bar R_\star}
\newcommand{\bRs}{\bar R_\star}
\newcommand{\Omst}{\Omega_\star}
\newcommand{\bOmst}{\bar \Omega_\star}
\newcommand{\kq}{k_{\rm q}}
\newcommand{\kst}{k_\star}
\newcommand{\ks}{k_\star}
\newcommand{\Msun}{\text{M}_\odot}
\newcommand{\Rsun}{\text{R}_\odot}

\newcommand{\Mp}{M_{\rm p}}
\newcommand{\ap}{a_{\rm p}}
\newcommand{\hp}{h_{\rm p}}
\newcommand{\bLp}{{\bm L}_{\rm p}}
\newcommand{\blp}{\bm {\hat l}_{\rm p}}
\newcommand{\Lp}{L_{\rm p}}
\newcommand{\ld}{\bm {\hat l}_{\rm d}}

\newcommand{\be}{\begin{equation}}
\newcommand{\ee}{\end{equation}}

\begin{document}


\title[Planets in Star-Disk-Binary Systems]
{Planet Formation in Disks with Inclined Binary Companions: Can Primordial Spin-Orbit Misalignment be Produced?}


\author[J. J. Zanazzi and Dong Lai]{J. J. Zanazzi$^{1}$\thanks{Email: jjz54@cornell.edu}, and Dong Lai$^{1}$ \\
$^{1}$Cornell Center for Astrophysics, Planetary Science, Department of Astronomy, Cornell University, Ithaca, NY 14853, USA}



\maketitle
\begin{abstract}
Many hot Jupiter (HJ) systems have been observed to have their stellar
spin axis misaligned with the planet's orbital angular momentum axis.  The
origin of this spin-orbit misalignment and the formation mechanism of
HJs remain poorly understood. A number of recent works have suggested
that gravitational interactions between host stars, protoplanetary disks, and
inclined binary companions may tilt the stellar spin axis with respect
to the disk's angular angular momentum axis, producing planetary
systems with misaligned orbits. These previous works considered
idealized disk evolution models and neglected the gravitational influence of newly formed planets.  In this paper, we explore how disk
photoevaporation and planet formation and migration affect the inclination evolution of
planet-star-disk-binary systems.  We take into account planet-disk interactions and the gravitational spin-orbit coupling between the host star and the planet.  We find that the rapid depletion of
the inner disk via photoevaporation reduces the excitation of stellar
obliquities. Depending on the formation and migration history of HJs,
the spin-orbit coupling between the star and the planet may reduces and even completely suppress the excitation
of stellar obliquities.
Our work constrains the formation/migration history of HJs. On the other hand, planetary systems with ``cold'' Jupiters or close-in super-earths may experience excitation of stellar obliquities in the presence of distant inclined companions.
\end{abstract}


\section{Introduction}
\label{sec:Intro}




Many exoplanetary systems containing hot Jupiters (HJs, giant planets
with periods of order a few days) have been found to have their orbital
angular momentum axis significantly misaligned with the spin axis of
the host star (e.g. \citealt{Hebrard(2008),Narita(2009),Winn(2009),Triaud(2010)}; see \citealt{WinnFabrycky(2015),Triaud(2017)} for recent reviews).
This ``spin-orbit misalignment'' is unexpected for a planet formed in
a protoplanetary disk, as a young star's spin axis is expected to be
aligned with the disk's angular momentum vector.  One explanation is HJs are formed through high-eccentricity channels, in which the
planet is pumped into a very eccentric orbit as a result of
gravitational interactions with other planets or with a distant
stellar companion, followed by tidal dissipation which circularizes
the planet's orbit (e.g., \citealt{WuMurray(2003),FabryckyTremaine(2007),Nagasawa(2008),WuLithwick(2011),Naoz(2012),BeaugeNesvorny(2012),Petrovich(2015),Anderson(2016),Munoz(2016),Hamers(2016)}). 
In this ``high-eccentricity migration'' scenario, the chaotic spin
evolution of the parent star driven by the changing orbit of the planet
(even for planets which do not suffer ``orbit flips'') plays the
dominant role in setting the final spin-orbit misalignment \citep{Storch(2014),StorchLai(2015),Storch(2017)}. Currently, it is unclear what
fraction of HJs are formed through these high-eccentricity routes,
and several observations remain difficult to explain, such as the
lack of giant planets with high eccentricities \citep{Dawson(2015)},
and the correlation between the spin-orbit misalignment and the
effective temperature of the host star (e.g., \citealt{Albrecht(2012),Mazeh(2015),LiWinn(2016),Winn(2017)}).


Other mechanisms have been proposed to explain spin-orbit
misalignments of HJ systems.  One idea is that the misalignment is
indicative of stellar astrophysics rather than planetary formation.
In \cite{Rogers(2012)}, it was suggested that internal gravity waves in
massive stars may transport angular momentum in the radiative
envelope, altering the star's surface rotation direction in a quasi-periodic manner.


There is observational evidence that a non-negligible fraction of HJs
may be formed in protoplanetary disks in-situ or through disk-driven migration.
For example, HJs (or hot Neptunes) around young T Tauri
stars have recently been detected \citep{Donati(2016),David(2016)}; 
such young HJs can only form in protoplanetary disks or
through disk-driven migration.  The HJ WASP-47b has two low-mass
neighbors \citep{Becker(2015)}, and thus cannot be formed through
high-eccentricity migration. \cite{Boley(2016)} and \cite{Batygin(2016)}
have advocated in-situ formation for such systems.  \cite{SchlaufmanWinn(2016)} found that HJs are equally likely to have exterior giant planet
companions inside the ice line compared to longer-period giant
planets, and argued against the high-$e$ migration scenario for HJ
formation. For HJs formed in-situ or through disk-driven migration, 
the observed stellar obliquities may result from 
``primordial misalignment," where spin-orbit misalignments are 
produced while the planets are embedded in the protoplanetary
disk.  Ways of generating primordially misaligned disks include chaotic
star formation \citep{Bate(2010),Fielding(2015)}, dynamical encounters
with other proto-stellar systems \citep{Thies(2011)}, magnetic star -
disk interactions \citep{Lai(2011),FoucartLai(2011)}, and gravitational interactions
with inclined planets \citep{MatsakosKonigl(2017)}.

\cite{Batygin(2012)} first suggested that the gravitational torque
from an inclined binary companion can change the orientation of a
protoplanetary disk with respect to its host star.
\cite{BatyginAdams(2013)}, \cite{Lai(2014)} and
\cite{SpaldingBatygin(2014)} included the gravitational coupling
between the host star and the disk, and showed that a secular resonance
occurs during the disk evolution, leading to a robust excitation of
misalignment between the stellar spin axis and the disk axis. Although
these works incorporated various effects such as stellar winds,
stellar contraction, accretion and magnetic star-disk interactions,
the disk physics included was highly idealized.  In particular, these previous 
works assumed a flat disk with homologous surface density evolution (i.e. the disk density profile remains constant in shape but decreases
in magnitude during the disk evolution).  Moreover, although these works
aimed at explaining the misalignment between the planet's orbit and
the spin of the host star, the gravitational influence of a massive
planet on the dynamics of the star-disk-binary system was neglected.


In this paper we study how the non-homologous surface density
evolution of disks due to photoevaporation and the formation/migration
of a planet orbiting close to its host star influence the generation
of spin-orbit misalignments in star-disk-binary systems.  In a
companion paper \citep{ZanazziLai(2017b)} we consider non-flat
(warped) disks and examine the effect of viscous dissipation from disk
warping on the spin-disk misalignments. Our paper is organized as
follows.  Section~\ref{sec:Review} reviews the physics of stellar
obliquity excitation through star-disk-binary interactions.
Section~\ref{sec:Photo} introduces a prescription parameterizing how
photoevaporation affects the disk's surface density evolution, and
studies how such evolution affects the inclination excitation in
star-disk-binary systems.  Section~\ref{sec:PlanetInts} presents an
overview of how an inclined planet interacts with the disk, the
central oblate star, and the distant binary companion.
Section~\ref{sec:DynPSDB} investigates how the formation/migration of
a short-period, massive planet affects the inclination evolution of
star-disk-binary systems.  We discuss the theoretical uncertainties and
observational implications in Section~\ref{sec:Discuss}, and provide a
summary of our key results in Section~\ref{sec:Conc}.

\section{Spin-Disk Misalignment from Star-Disk-Binary Gravitational Interactions}
\label{sec:Review}
Previous works \citep{BatyginAdams(2013),Lai(2014),SpaldingBatygin(2014)} have shown that secular resonance can generate misalignment between the stellar spin and protoplanetary disk in star-disk-binary systems.  In this section, we set up the problem and review the main physics behind this mechanism.  We assume the disk is flat with orbital angular momentum unit vector $\ld$, justified in a companion paper \citep{ZanazziLai(2017b)}.  Our treatment follows \cite{Lai(2014)} (hereafter L14) based on the dynamics of angular momentum vectors.  For clarity, we display all quantities defined in Sections~\ref{sec:Review}-\ref{sec:Photo} in Table~\ref{table:Table1}.

\begin{table}
\begin{tabular}{| l | l | l |}
\hline
Symbol & Meaning & Eq. \\
\hline
$\Mst$ & central star mass & - \\
$\Rst$ & central star radius & - \\
$\Omst$ & central star's rotation rate & - \\
$P_\star$ & central star's rotation period & - \\
$\Md$ & disk mass & \eqref{eq:Md} \\
$\rin$ & disk inner truncation radius & - \\
$\rout$ & disk outer truncation radius & - \\
$\Mb$ & binary mass & - \\
$\ab$ & binary semi-major axis & - \\
$\bar X$ & normalized quantity $X$ & \eqref{eq:pars}\\
$\Sg$ & disk surface density & - \\
$\Sgout$ & disk surface density at $r = \rout$ & \eqref{eq:p} \\
$p$ & power-law surface density index & \eqref{eq:p} \\
$\Ld$ & disk total orbital angular momentum & \eqref{eq:Ld} \\
$S$ & stellar spin angular momentum & \eqref{eq:S} \\
$k_\star$ & stellar spin normalization & \eqref{eq:S} \\
$\kq$ & stellar quadrupole moment normalization & - \\
$\bs$ & stellar spin unit vector & - \\
$\ld$ & disk orbital angular momentum unit vector & - \\
$\bomds$ & precession rate of disk around star & \eqref{eq:bomds} \\
$\bomsd$ & precession rate of star around disk & \eqref{eq:bomsd} \\
$\bomdb$ & precession rate of disk around binary & \eqref{eq:bomdb} \\
$M_{\rm d0}$ & disk initial mass &- \\
$\tv$ & disk viscous timescale & \eqref{eq:Md_evolv} \\
${\bar M}_{\rm d0}$ & normalized disk initial mass & - \\
$\tgsd$ & mutual star-disk inclination & \eqref{eq:tgsd} \\
$\tgsb$ & mutual star-binary inclination & \eqref{eq:tgsb} \\
$\tgdb$ & mutual disk-binary inclination & \eqref{eq:tgdb} \\
$\rc$ & critical photoevaporation radius & - \\
$\tw$ & critical photoevaporation time & - \\
$t_{\rm v,out}$ & outer disk's viscous time & - \\
$t_{\rm v,in}$ & inner disk's viscous time & - \\
$\Sgc$ & disk surface density at $r = \rc$ & - \\
$\bomsdg$ & precession rate of star & {} \\
{} & around disk exterior to $\rc$ & \eqref{eq:bomsdg} \\
$\bomsdl$ & precession rate of star & {} \\
{} & around disk interior to $\rc$ & \eqref{eq:bomsdl} \\
$\bomdgs$ & precession rate of disk & {} \\
{} & interior to $\rc$ around star & \eqref{eq:bomdgs} \\
$\bomdls$ & precession rate of disk & {} \\
{} & exterior to $\rc$ around star & \eqref{eq:bomdls}
\end{tabular}
\caption{Definitions of relevant quantities in the star-disk-binary system.}
\label{table:Table1}
\end{table}

\subsection{Setup and Parameters}
\label{sec:Setup}

Consider a central star of mass $\Mst$, radius $\Rst$, rotation rate $\Omst$, with a circumstellar disk of mass $\Md$, and inner and outer truncation radii of $\rin$ and $\rout$, respectively.  This star-disk system is in orbit with a distant binary companion of mass $\Mb$ and semimajor axis $\ab$.  We introduce the following rescaled parameters typical of protostellar systems:
\begin{align}
&\bMst = \frac{\Mst}{1 \, \Msun},
\hspace{3mm}
\bRst = \frac{\Rst}{2 \, \Rsun},
\hspace{3mm}
\bOmst = \frac{\Omst}{\sqrt{G \Mst/\Rst^3}},
\nonumber \\
&\bMd = \frac{\Md}{0.01 \, \Msun},
\hspace{3mm}
\brin = \frac{\rin}{8 \, \Rsun},
\hspace{3mm}
\brout = \frac{\rout}{50 \, \text{au}},
\nonumber \\
&\bMb = \frac{\Mb}{1 \, \Msun},
\hspace{5mm}
\bab = \frac{\ab}{300 \, \text{au}}.
\label{eq:pars}
\end{align}
The cannonical value of $\bOmst$ is $0.1$, corresponding to a stellar rotation period of $P_\star = 3.3 \, \text{days}$.  The other canonical values in Eq.~\eqref{eq:pars} are unity, except the disk mass, which can change significantly during the disk lifetime.

In the simplest model, we parameterize the disk surface density $\Sg = \Sg(r,t)$ as
\be
\Sg(r,t) = \Sgout(t) \left( \frac{\rout}{r} \right)^p.
\label{eq:p} 
\ee
L14 used $p=1$.  We will introduce a more complex parameterization of $\Sg(r,t)$ in Section~\ref{sec:Photo} to account for the effect of photoevaporation.  We choose $p$ between $1$ and $3/2$.  This choice is motivated by various observations.  In the outer regions of disks around YSO's ($ r \gtrsim \text{few} \, \text{au}$), $p$ is constrained to lie in between $\sim 0.5-1$ \citep{WilliamsCieza(2011)}.  For the inner regions ($r \lesssim \text{few} \, \text{au}$), direct observational constraints are lacking.  The Minimum Mass Solar Nebulae has $p = 3/2$ \citep{Weidenschilling(1977)}, and the Minimum Mass Extra-Solar Nebulae (assuming the planets discovered by \textit{Kepler} formed in-situ; \citealt{ChaingLaughlin(2013)}) have $p \simeq 1.6$.   The main effect of increasing $p$ is to increase the amount of mass available to form a short-period gas-giant planet [see Eq.~\eqref{eq:Mp}], and increase the mutual star-disk precession frequencies [see Eqs.~\eqref{eq:bomds}-\eqref{eq:bomsd}].  We will always assume $p < 2$ when calculating global disk properties (mass, angular momentum, precession frequencies, etc.), the expressions for many of these quantities will differ when $p \ge 2$ in the limit $\rin \ll \rout$.

The disk mass $\Md$ is then (assuming $\rin \ll \rout$)
\be
\Md = \int_{\rin}^{\rout} 2 \pi \Sg r \der r \simeq \frac{2\pi \Sgout \rout^2}{2-p} .
\label{eq:Md}
\ee
The disk angular momentum vector is $\bLd = \Ld \ld$, and the stellar spin angular momentum vector is $\bS = S \bs$, where $\ld$ and $\bs$ are unit vectors, and
\begin{align}
\Ld &= \int_{\rin}^{\rout} 2 \pi \Sg r^3 \Om \der r \simeq \frac{2-p}{5/2-p} \Md  \sqrt{G \Mst \rout},
\label{eq:Ld} \\
S &= k_\star \Mst \Rst^2 \Omst,
\label{eq:S}
\end{align}
with $\Om(r) \simeq \sqrt{G \Mst/r^3}$ and $k_\star \simeq 0.2$.

\subsection{Gravitational Torques}
\label{sec:GravTorques}

The stellar rotation leads to a difference in the principal components of the star's moment of inertia of $I_3 - I_1 = \kq \Mst \Rst^2 \bOmst^2$, where $\kq \simeq 0.1$ for fully convective stars \citep{Lai(1993)}.  The gravitational torque on the disk from the star is\footnote{Throughout this paper, quantities with a tilde ($\tilde {\ }$) imply an average or integration over the disk}
\begin{align}
\bbT_{\rm ds} &= -\int_{\rin}^{\rout} \frac{3 G (I_3 - I_1)}{2 r^3} (\ld \bcdot \bs) (\bs \btimes \ld) 2 \pi \Sg r^3 \Om \der r
\nonumber \\
&= - \Ld \bomds (\ld \bcdot \bs) \bs \btimes \ld,
\label{eq:Tds}
\end{align}
where (assuming $\rin \ll \rout$)
\be
\bomds \simeq \frac{3(5/2-p)\kq}{2(1+p)} \frac{\Rs^2 \bOmst^2}{\rout^{1-p} \rin^{1+p}} \sqrt{ \frac{G \Ms}{\rout^3} }
\label{eq:bomds}
\ee
characterizes the precession frequency of the disk around the star.  The back-reaction torque on the star from the disk is $\bbT_{\rm sd} = - \bbT_{\rm ds}$, and causes the star to precess around the disk at a characteristic frequency
\begin{align}
\bomsd &= ( \Ld / S ) \bomds
\nonumber \\
&\simeq \frac{3(2-p)\kq}{2(1+p)\ks} \left( \frac{\Md}{\Ms} \right) \bOmst \frac{\sqrt{G \Ms \Rs^3}}{\rout^{2-p} \rin^{1+p}}.
\label{eq:bomsd}
\end{align}

The torque on the disk from the inclined binary companion is (assuming $\rout \ll \ab$)
\begin{align}
\bbT_{\rm db} &\simeq - \int_{\rin}^{\rout} \left( \frac{3 G \Mb r^2}{4 \ab^3} \right) (\ld \bcdot \blb)(\blb \btimes \ld) 2 \pi \Sg r^3 \Om \der r
\nonumber \\
&= -\Ld \bomdb (\ld \bcdot \blb) \blb \btimes \ld,
\label{eq:Tdb}
\end{align}
where
\begin{align}
\bomb \simeq \frac{3(5/2-p)}{4(4-p)} \left( \frac{\Mb}{\Ms} \right) \left( \frac{\rout}{\ab} \right)^3 \sqrt{ \frac{G \Ms}{\rout^3} }
\label{eq:bomdb}
\end{align}
characterizes the precession frequency of the disk around the binary.

Taking $p = 1$, the precession frequencies~\eqref{eq:bomds}, \eqref{eq:bomsd}, and~\eqref{eq:bomdb} evaluate to
\begin{align}
\boms &= 2.0 \times 10^{-7} \left( \frac{\kq}{0.1} \right) \frac{\bRst^2 \bMst^{1/2}}{\brin^2 \brout^{3/2}} \left( \frac{\bOmst}{0.1} \right)^2 \left( \frac{2\pi}{\text{yr}} \right),
\label{eq:boms_num} \\
\bomb &= 4.9 \times 10^{-6} \frac{\bMb \brout^{3/2} }{\bMst^{1/2} \bab^3} \left( \frac{2\pi}{\text{yr}} \right),
\label{eq:bomb_num} \\
\bomd &= 4.9 \times 10^{-5} \left( \frac{ 2 \kq}{\kst} \right) \frac{\bMd}{\bMst^{1/2} \bRst^{1/2} \brin^2 \brout} \left( \frac{\bOmst}{0.1} \right) \left( \frac{2\pi}{\text{yr}} \right).
\label{eq:bomd_num}
\end{align}
\subsection{System Evolution and Secular Resonance}
\label{sec:SecRes}

\begin{figure}
\centering
\includegraphics[scale=0.6]{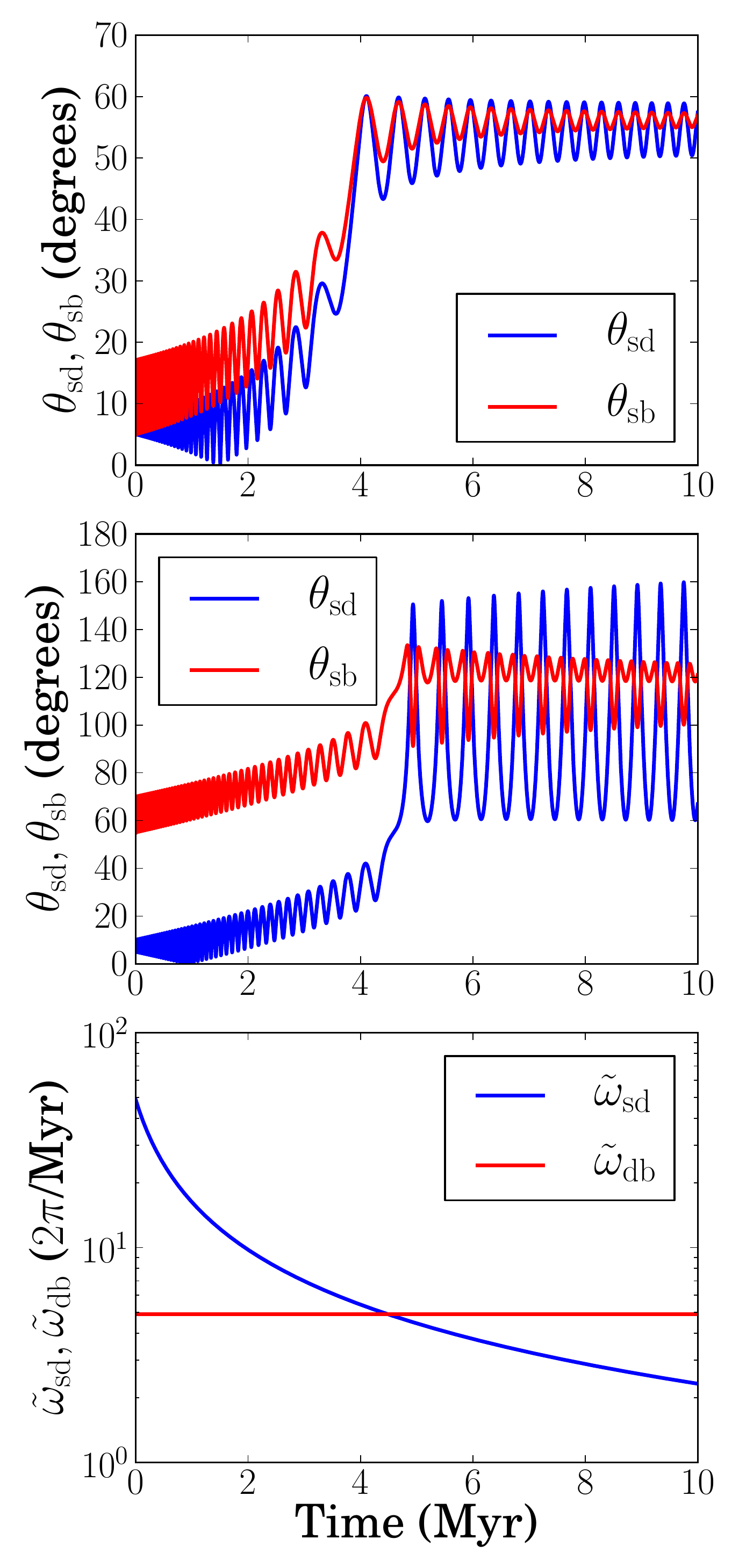}
\caption{
Sample evolution of the star-disk-binary system.  The top and middle panels plot the evolution of the star-disk inclination $\tgsd$ [Eq.~\eqref{eq:tgsd}] and star-binary inclination $\tgsb$ [Eq.~\eqref{eq:tgsb}].  We take the initial disk-binary inclination $\tgdb$ to be $\tgdb(0) = 10^\circ$ (top panel) and $\tgdb(0) = 60^\circ$ (middle panel) with $\tgsd(0) = 5^\circ$ for both panels.  The bottom panel plots the time evolution of the precession rates $\bomsd$ [Eq.~\eqref{eq:bomsd}] and $\bomdb$ [Eq.~\eqref{eq:bomdb}].    We take all parameter values to be canonical~[Eq.~\eqref{eq:pars}].
}
\label{fig:tg_stand}
\end{figure}

The time evolution of the star-disk-binary system is given by
\begin{align}
\frac{\der \bs}{\der t}  &= - \bomsd(\bs \bcdot \ld) \ld \btimes \bs,
\label{eq:dsdt_grav} \\
 \frac{\der \ld}{\der t}  &=  - \bomds (\ld \bcdot \bs) \bs \btimes \ld - \bomdb (\ld \bcdot \blb)\blb \btimes \ld.
\label{eq:dldt_grav}
\end{align}
As in \cite{BatyginAdams(2013)} and \cite{Lai(2014)}, we assume the disk mass evolves according to
\be
\Md = \frac{M_{ {\rm d}0}}{1+ t/\tv}.
\label{eq:Md_evolv}
\ee
For our canonical parameters, we choose $M_{{\rm d}0} = 0.1 \, M_\odot$ and $\tv = 0.5 \, \text{Myr}$.  We define $\bar M_{\rm d0} = M_{\rm d0} / 0.1 \, \Msun$.

Figure~\ref{fig:tg_stand} shows an example of the star-disk-binary system evolution.  We define the angles
\begin{align}
\tgsd &= \cos^{-1}(\bs \bcdot \ld), 
\label{eq:tgsd} \\
\tgsb &= \cos^{-1}(\bs \bcdot \blb),
\label{eq:tgsb} \\
\tgdb &= \cos^{-1}(\ld \bcdot \blb).
\label{eq:tgdb}
\end{align}
The angles $\tgsd$, $\tgsb$, and $\tgdb$ denote the mutual star-disk, star-binary, and disk-binary inclinations, respectively. We take $\bs$, $\bl$, and $\blb$ to initially all lie in the same plane, with $\tg_{\rm sd}(0) = 5^\circ$ and two different values of $\tg_{\rm db}(0)$. We choose our initial value of $\tgsd$  to be $\tgsd(0) \ll 1$.  The dynamics of the star-disk-binary system remain qualitatively unchanged as long as $\tgsd(0)$ is much smaller than unity \citep{SpaldingBatygin(2014)}.

When $\bomb \ll \bomd$ early in the disk's lifetime, $\bs$ adiabatically tracks $\ld$, and $\tgsd \sim \text{constant}$.  When $\bomb \gg \bomd$ later in the disk's lifetime, $\bs$ tracks $\blb$ with $\tgsb \sim \text{constant}$.  A secular resonance occurs when $\bomb \sim \bomd$, and large $\tgsd$ can be generated due to the change in the dynamical behavior of the stellar spin axis (see L14 for discussion).  This resonant excitation of $\tgsd$ is prominent when $\Ld \gtrsim S$ at the resonance crossing.

\section{Non-homologous Surface Density Evolution: Photoevaporation}
\label{sec:Photo}

Section \ref{sec:Review} assumes (as in previous works) the disk surface density evolves homologously, maintaining the power-law $r^{-p}$ profile while decreasing in the overall magnitude. Realistic protostellar disks do not evolve in such a homologeous way.  This section explores an alternate prescription for the surface density evolution that captures the essential physics of photoevaporation (e.g. \citealt{Clarke(2001),Alexander(2014)}).

\begin{figure}
\centering
\includegraphics[scale=0.4]{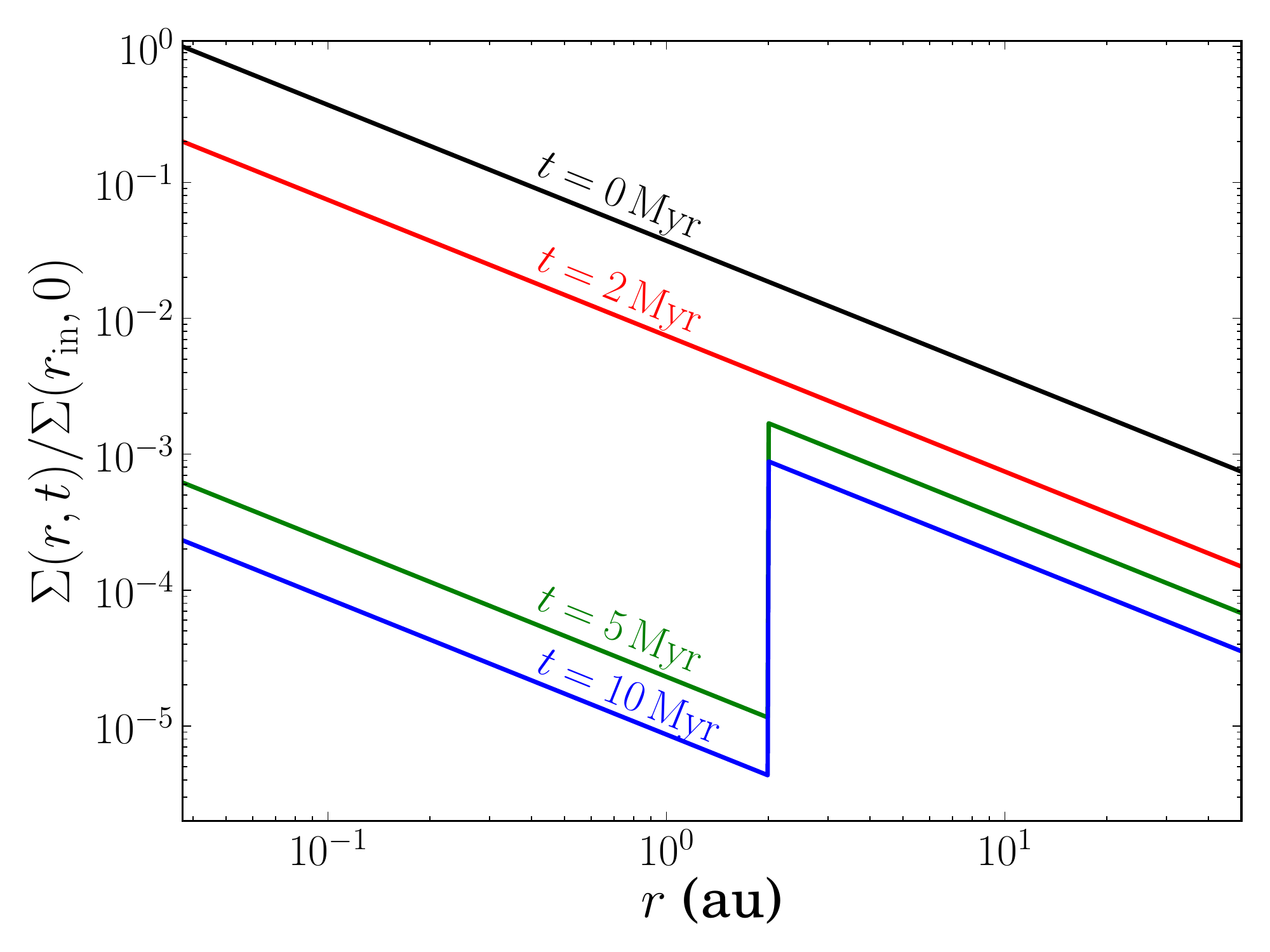}
\caption{
Evolution of the disk surface density $\Sg(r,t)$, given by Eq.~\eqref{eq:Sg_photo}.  We take  the disk's outer viscous timescale to be $\tv = 0.5 \, \text{Myr}$, the time when the disk's photo-ionization rate is comparable to viscous depletion rate $\tw = 2 \, \text{Myr}$, the disk's inner viscous time $\tvin = 0.02 \, \text{Myr}$, and the critical radius separating the inner and outer regions of the disk $\rc = 2 \, \text{au}$.
}
\label{fig:Sg_photo}
\end{figure}

\begin{figure}
\centering
\includegraphics[scale=0.6]{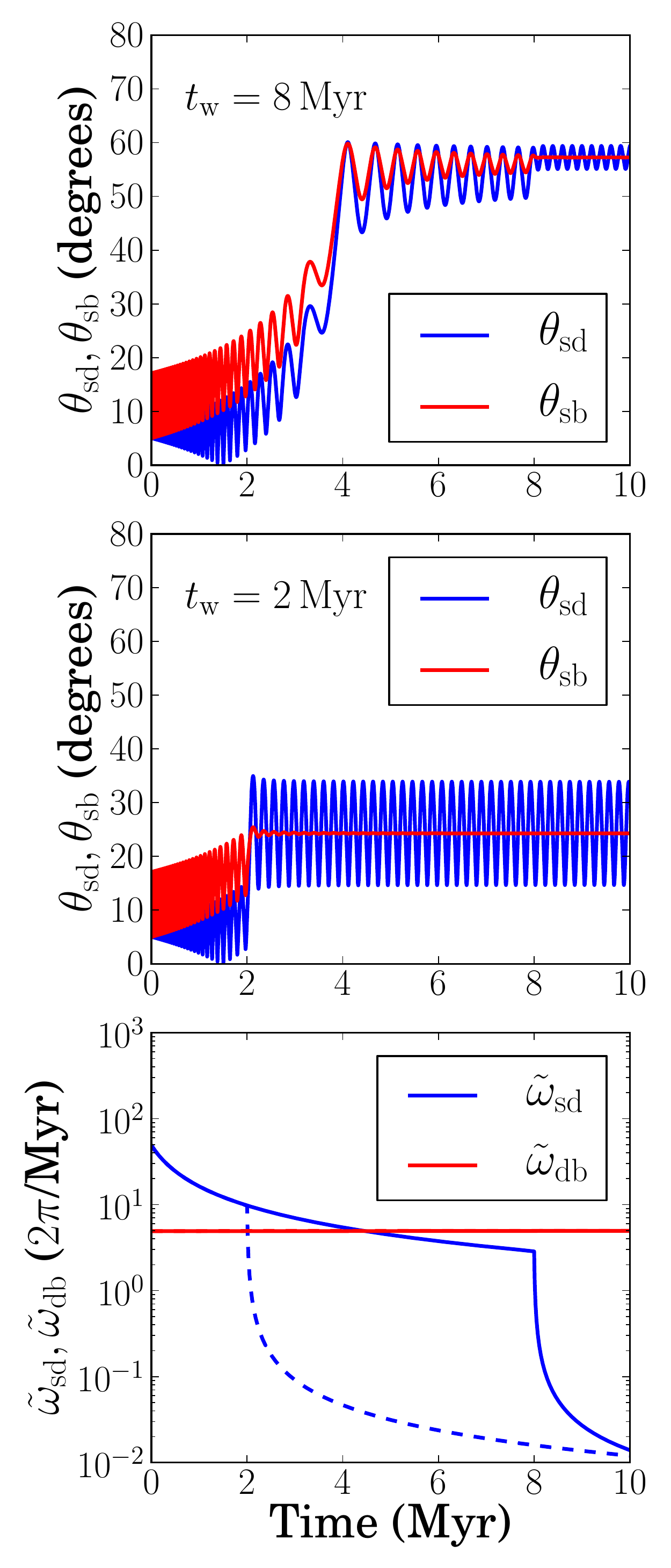}
\caption{
Same as Fig.~\ref{fig:tg_stand}, except the disk surface density $\Sg(r,t)$ evolves according to the prescription~\eqref{eq:Sg_photo}, with values of $\tw$ as indicated.  The bottom panel shows the  disk-binary precession rate $\bomb$ [Eq.~\eqref{eq:bomdb}] and the star-disk precession rate $\bomd$ [Eq.~\eqref{eq:bomsd_photo}], with $\tw = 8 \, \text{Myr}$ (solid), and $\tw = 2 \, \text{Myr}$ (dashed).  We take $\rc = 2 \, \text{AU}$, $\tv = 0.5 \, \text{Myr}$, $\tvin = 0.02 \, \text{Myr}$.  All other parameters canonical, with the initial star-disk inclination $\tgsd(0) = 5^\circ$ and disk-binary inclination $\tgdb(0) = 10^\circ$. 
}
\label{fig:tg_photo}
\end{figure}

As described in \cite{Clarke(2001)}, the combined influence of photoevaporation and viscous accretion dramatically influence the surface density evolution of disks around T-Tauri stars (see \citealt{Alexander(2014),Owen(2016)} for recent reviews).  The surface density $\Sg$ has distinct behaviors before and after the characteristic time  $\tw$, when the viscous accretion rate and photo-evaporative mass loss rate become comparable at the critical photoevaporation radius $r_{\rm c} \sim \text{a few au}$ (the maximal radius where photoionized gas remains bound to the central star; \citealt{Hollenbach(1994),Alexander(2006)}).  Before $\tw$, viscous accretion drives the disk's mass depletion, and the surface density evolves over the outer disk's viscous time $\tvout = \tv$.  After $\tw$, $\Sg$ at $r > r_{\rm c}$ continues to evolve viscously over the timescale $\tvout = \tv$ [see Eq.~\eqref{eq:Md_evolv}]; interior to $\rc$, photoevaporation starves the inner disk from resupply by the outer disk's viscous evolution, and the inner disk is drained over the inner disk's viscous time $\tvin \ll \tvout$.

To capture the main effect of photoevaporation, we parameterize the disk's surface density evolution as
\be
\Sg(r,t) = \left\{
\begin{array}{cc}
\Sg_{\rm c}(t)(\rc/r)^p & \rin \le r \le \rc \\
\Sg_{\rm out}(t)(\rout/r)^p & \rc < r \le \rout
\end{array}
\right.
\label{eq:Sg_photo},
\ee
where $\Sg_{\rm out}(t) = \Sgout(0)/(1+t/\tv)$ [see Eq.~\eqref{eq:Md_evolv}], while
\be
\Sgc(t) = \left\{
\begin{array}{cc}
\Sgc(0)(1+t/\tv)^{-1} & t \le \tw \\
\Sgc(\tw)[1+(t-\tw)/\tvin]^{-1} & t > \tw
\end{array}
\right.
\label{eq:Sgc(t)},
\ee
and
\be
\Sgc(0) =  \Sgout(0) ( \rc/\rout )^p.
\ee
Figure~\ref{fig:Sg_photo} shows a sample evolution of $\Sg(r,t)$.  Our prescription of $\Sg(r,t)$ introduces three new parameters: $\tw$ (when the inner disk begins to be rapidly depleted), $t_{\rm v,in}$ (the timescale over which the inner disk is depleted), and $\rc$ (the critical radius separating the inner and outer disks).  Observations constrain $\tw \sim 10^6 - 10^7 \, \text{years}$, $t_{\rm v,in} \sim 10^2 - 10^5 \, \text{years}$, and $\rc \sim \text{few} \times \text{au}$ \citep{Alexander(2014),Owen(2016)}.  We choose $\rc = 2 \, \text{au}$ throughout this paper, varying $\tw$ and $t_{\rm v,in}$ for different disk models.

To model the dynamics of the star-disk-binary system, we neglect any misalignments which may develop between the inner and outer disk planes, since the gravitational influence of the inner disk quickly becomes irrelevant to the dynamics of the star-disk-binary system.  Coplanarity between these two disk planes is maintained via bending waves \citep{PapaloizouLin(1995),LubowOgilvie(2000),ZanazziLai(2017b)} and disk self-gravity \citep{Batygin(2012),BatyginAdams(2013),ZanazziLai(2017a)}.  

The modified surface density evolution alters the mutual star-disk precession frequencies:
\begin{align}
\bomsd &= \bomsdl + \bomsdg,
\label{eq:bomsd_photo} \\
\bomds &= \bomdls + \bomdgs.
\label{eq:bomds_photo}
\end{align}
Here, ${\rm d<}$ (${\rm d>}$) denotes the disk interior (exterior) to $\rc$.  In terms of model parameters, the frequencies in Eqs.~\eqref{eq:bomsd_photo}-\eqref{eq:bomds_photo} evaluate to be (assuming $\rin \ll \rc \ll \rout$)
\begin{align}
\bomsdg &\simeq \frac{3 \kq}{(1+p)\ks} \bOmst \left( \frac{\pi \Sgout \rout^2}{\Ms} \right) \frac{ \sqrt{G \Ms \Rs^3} }{\rc^{1+p} \rout^{2-p}},
\label{eq:bomsdg} \\
\bomsdl &\simeq \frac{3 \kq}{(1+p)\ks} \bOmst \left( \frac{\pi \Sgc \rc^2}{\Ms} \right) \frac{ \sqrt{G \Ms \Rs^3} }{\rin^{1+p} \rc^{2-p}},
\label{eq:bomsdl} \\
\bomdgs &\simeq \frac{3(5/2-p)\kq}{2(1+p)} \bOmst^2 \frac{\Rs^2}{\rc^{1+p}\rout^{1-p}} \sqrt{ \frac{G \Ms}{\rout^3} },
\label{eq:bomdgs} \\
\bomdls &\simeq \frac{3(5/2-p)\kq}{2(1+p)} \bOmst^2 \frac{\Sgc \rc^p \Rs^2}{\Sgout \rin^{1+p} \rout} \sqrt{ \frac{G \Ms}{\rout^3} }.
\label{eq:bomdls}
\end{align}
The disk-binary precession frequency [$\bomdb$, Eq.~\eqref{eq:bomdb}] is unchanged (assuming $\rc \ll \rout$). The frequency $\bomsdg$ ($\bomsdl$) denotes the precession frequency of the star around the disk exterior (interior) to $\rc$, while $\bomdgs$ ($\bomdls$) denotes the precession frequency of the disk exterior (interior) to $\rc$ around the star.

Figure~\ref{fig:tg_photo} shows examples of the star-disk-binary evolution under the $\Sg(r,t)$ prescription~\eqref{eq:Sg_photo}, for two values of $\tw$.  We see that the main effect of photoevaporation is a potential change in resonance crossing time.  If the resonance ($\bomd \sim \bomb$) occurs before $\tw$, the excitation of $\tgsd$ is more or less unaffected.  If the resonance occurs after $\tw$, the rapid depletion of the inner disk causes $\bomd$ to rapidly approach zero over the time $\tvin$, and $\bomd \sim \bomb$ at $t \approx \tw+\tvin$.  The resuting $\tgsd$ excitation is smaller because the resonance crossing is fast. In either case, after $\tw$, the spin-binary misalignment angle $\tgsb$ freezes to a constant value because of the greatly diminished inner disk mass.

\section{Planet-Star-Disk-Binary Interactions}
\label{sec:PlanetInts}

We now add a planet in our star-disk-binary sytem.  This section examines how the planet interacts with the protoplanetary disk, the host star, and the inclined binary.  We take the planet to lie on a circular orbit, with mass $\Mp$, semi-major axis $\ap$, and orbital angular momentum $\bLp = \Mp \sqrt{G \Ms \ap} \blp$.  For clarity, we display all quantities defined in Sections~\ref{sec:PlanetInts}-\ref{sec:DynPSDB} in Table~\ref{table:Table2}.

\begin{table}
\begin{tabular}{| l | l | l |}
\hline
Symbol & Meaning & Eq. \\
\hline
$\Mp$ & planet mass & - \\
$\ap$ & planet semi-major axis & - \\
$\blp$ & planet orbital angular momentum unit vector & - \\
$h$ & disk aspect ratio & - \\
$\Dg_{\rm p}$ & gap width & - \\
$\bar \Sg$ & average disk surface density at gap edges & - \\
$\hp$ & maximum of $h(\ap)$ and $\Dg_{\rm p}/\ap$ & - \\
$\bompd$ & precession rate of planet around disk & \eqref{eq:bompd} \\
$\bomdb$ & precession rate of disk around planet & \eqref{eq:bomdp} \\
$\Lambda_{\rm mig}$ & Migration rate free parameter & \eqref{eq:t_mig} \\
$t_{\rm mig}$ & type II migration timescale & \eqref{eq:t_mig} \\
$\bompdl$ & precession rate of planet& {} \\
{} & around disk interior to $\rc$ & \eqref{eq:bompdl} \\
$\bompdg$ & precession rate of planet& {}\\
{} &around disk exterior to $\rc$ & \eqref{eq:bompdg} \\
$\bomdlp$ & precession rate of disk & {} \\
{} & interior to $\rc$ around planet & \eqref{eq:bomdlp} \\
$\bomdgp$ & precession rate of disk& {} \\
{} & exterior to $\rc$ around planet & \eqref{eq:bomdgp} \\
$\omps$ & precession rate of planet around star & \eqref{eq:omps} \\
$\omsp$ & precession rate of star around planet & \eqref{eq:omsp} \\
$\ompb$ & precession rate of planet around binary & \eqref{eq:ompb} \\
$\xi$ & feeding zone free parameter & \eqref{eq:Dgap} \\
$\tg_{\rm ps}$ & mutual planet-star inclination & \eqref{eq:tgps} \\
$\tg_{\rm pd}$ & mutual planet-disk inclination & \eqref{eq:tgpd} \\
$\tg_{\rm pb}$ & mutual planet-binary inclination & \eqref{eq:tgpb} \\
$\bomsd'$ & modified precession frequency & {} \\
{} & of star around disk & \eqref{eq:bomsd_p} \\
$\bomds'$ & modified precession frequency & {} \\
{} & of disk around star & \eqref{eq:bomds_p} \\
$\bomdb'$ & modified precession frequency & {} \\
{} & of disk around binary & \eqref{eq:bomdb_p}
\end{tabular}
\caption{Definitions of quantities related to planet interactions with the star-disk-binary system.}
\label{table:Table2}
\end{table}

\subsection{Planet-Disk Interactions: Non-Gap Opening Planets}
\label{sec:PlanetDiskNG}

  When the planet has a mass insufficient to open a gap in the disk, the gravitational torque on the planet from the disk causes $\blp$ to precess around $\ld$ at a rate \citep{Ward(1981),Hahn(2003)}
\be
\tilde \om_{\rm pd,\, no \, gap} \simeq \frac{\pi \Sg(\ap,t) \ap^2}{\Ms h(\ap)} \sqrt{ \frac{G \Ms}{\ap^3} }
\label{eq:bompd_ng}
\ee
where $\Sg(\ap,t)$ is the disk surface density at $r = \ap$, and $h(\ap)$ is the disk aspect ratio $h = H/r$ ($H$ is the disk scaleheight) evaluated at $r = \ap$.  Equation~\eqref{eq:bompd_ng} was derived assuming $|\blp \btimes \ld| \ll h(\ap) \ll 1$, using a disk potential with softening length $H$.  

In addition to the direct gravitational torque, when $\blp$ is misaligned with $\ld$, the planet drives bending waves which propagate through the disk, working to cause $\blp$ to precess and align with $\ld$ on a characteristic timescale \citep{TanakaWard(2004),Cresswell(2007),KleyNelson(2012)} 
\be
t_{\rm bw} = \frac{\Ms}{\Sg(\ap,t) \ap^2} \left( \frac{\Ms}{\Mp} \right) h^4(\ap) \sqrt{ \frac{\ap^3}{G \Ms} }.
\label{eq:tbw}
\ee
Since
\be
t_{\rm bw}\tilde \om_{\rm pd, \, no \, gap} = 1.05 \times 10^{3} \left( \frac{\Mp}{1 \, \ME} \right)^{-1} \bar M_\star \left( \frac{h(\ap)}{0.1} \right)^{3},
\label{eq:tbw_ng}
\ee
we expect $\blp$ to precess around $\ld$ mainly due to the gravitational torque, with the bending waves aligning $\blp$ with $\ld$ over a longer time-scale.

The planet also drives density waves in the disk, leading to its radial migration \citep{GoldreichTremaine(1979),Tanaka(2002),KleyNelson(2012)}.  The characteristic migration time is
\begin{align}
&t_{\rm mig, \, no \, gap} = \frac{t_{\rm bw}}{h^2(\ap)}
\nonumber \\
&= \frac{\Ms}{\Sg(\ap,t) \ap^2} \left( \frac{\Ms}{\Mp} \right) h^2(\ap) \sqrt{ \frac{\ap^3}{G \Ms} }.
\label{eq:t_mig_ng}
\end{align}
 The migration rate depends on the detailed local properties of the disk, such as if the disk lies in a dead zone (e.g. \citealt{McNally(2017)}), the local temperature gradient (e.g. \citealt{JimenezMasset(2017)}), and the disk's thermal diffusivity and planet's accretion rate \citep{Benitez-Llambay(2015),MassetVelascoRomero(2017),Masset(2017)}.  These effects may drive $\ap$ to increase or decrease with time.  Because the planet becomes dynamically important only when its mass becomes sufficiently large to open a gap, we will neglect its orbital evolution, and fix $\ap$ in time before a gap is opened.

\subsection{Planet-Disk Interactions: Gap Opening Planets}
\label{sec:PlanetDiskGap}

When the planet has sufficient mass $\Mp \gtrsim 40 \Ms \ag h^2(\ap)$ (e.g. \citealt{LinPapaloizou(1993)}), it can open a gap in the disk, with a width $\Dg_{\rm p} = \xi \ap (\Mp/3\Ms)^{1/3}$ ($\xi$ is a free parameter).  The disk surface density around the gap is clearly complex.  In our calculation, we adopt the simple prescription that $\Sg(r,t) \simeq 0$ for $|r - \ap| < \Dg_{\rm p}/2$, and $\Sg(r,t)$ obeys Eqs.~\eqref{eq:p} or~\eqref{eq:Sg_photo} otherwise.  The mutual planet-disk interactions are modified from the non-gap opening planet case.  When the planet opens a gap, because we expect the disk gravitational potential $\Phi_{\rm d}$ to not exceed $2 \pi G\bar \Sg(\ap,t) \ap^2/\Dg_{\rm p}$,  where $\bar \Sg(\ap,t) = \frac{1}{2} [\Sg(\ap - \Dg_{\rm p}/2,t) + \Sg(\ap + \Dg/2,t)]$, we may replace the softening length in the disk potential $\Phi_{\rm d}$ by $\Dg_{\rm p}$.  The characteristic precession frequency of the planet around the disk is then modified to become [cf. Eq.~\eqref{eq:bompd_ng}]
\be
\tilde \om_{\rm pd, \, gap} \simeq \frac{\pi \bar \Sg(\ap, t) \ap^3}{\Ms \Dg_{\rm p}} \sqrt{ \frac{G \Ms}{\ap^3} }.
\label{eq:bompd_gap}
\ee
 From now on, we define
\be
\bompd = \frac{\pi \bar \Sg(\ap, t) \ap^2}{\Ms h_{\rm p}} \sqrt{ \frac{G \Ms}{\ap^3} },
\label{eq:bompd}
\ee
where $h_{\rm p} = \max[h(\ap), \Dg_{\rm p}/\ap]$.  The planet exerts a back-reaction torque on the disk, causing $\ld$ to precess around $\blp$ at a characteristic rate
\begin{align}
\bomdp &= (\Lp/\Ld)\bompd
\nonumber \\
&\simeq \frac{5/2 - p}{2 h_{\rm p}} \left( \frac{ \ap}{\rout} \right)^{1-p} \left( \frac{\Mp}{\Ms} \right) \sqrt{ \frac{G \Ms}{\rout^3} }.
\label{eq:bomdp}
\end{align}

Because bending waves propagate through the disk as a result of resonant Lindblad and co-rotational torques, a gap $\Dg_{\rm p}$ lengthens $t_{\rm bw}$ to be [c.f. Eq.~\eqref{eq:tbw_ng}]
\be
t_{\rm bw} = \Lam_{\rm gap}\frac{\Ms}{\bar \Sg(\ap) \ap^2} \left( \frac{\Ms}{\Mp} \right) h^4(\ap) \sqrt{ \frac{\ap^3}{G \Ms} }.
\label{eq:tbw_gap}
\ee
The numerical value of $\Lam_{\rm gap}$ must be obtained via hydrodynamical simulations to account for non-linear effects.  No simulations have carefully calculated $\Lam_{\rm gap}$ as a function of the planet's parameters and local disk properties, but simulations suggest $\Lam_{\rm gap} \gg 1$ (e.g. \citealt{Xiang-GruessPapaloizou(2013),Bitsch(2013),Chametla(2017)}).  Comparing $\bompd$ to $t_{\rm bw}$,
\begin{align}
\tilde \om_{\rm pd, \, gap} t_{\rm bw} = & \ 3.29 \times 10^2 \left( \frac{\Lam_{\rm gap}}{100} \right) \left( \frac{\Mp}{1 \, \MJ} \right)^{-1}
\nonumber \\
&\times \bar M_\star \left( \frac{h(\ap)}{0.1} \right)^{4} \left( \frac{\Dg_{\rm p}/\ap}{0.1} \right)^{-1}.
\end{align}
From this, we see a gap-opening planet should interact with a disk mainly through gravitational torques.

When a planet opens a gap in the disk, $\ap$ evolves due to the disk's viscous evolution (Type II migration).  If $\Mp \lesssim 2\pi \bar \Sg(\ap) \ap^2$, the planet follows the viscous evolution of the disk, and $\ap$ decreases over the disk's viscous time $\tv$ \citep{LinPapaloizou(1985),Lin(1996),KleyNelson(2012)}.  When $\Mp \gtrsim 2\pi \bar \Sg(\ap) \ap^2$, the planet's gravitational torque balances the disk's viscous torque, and migrates inward over a timescale longer than $\tv$ \citep{LinPapaloizou(1985),IdaLin(2004),KleyNelson(2012)}. Motivated by simulations of gap-opening planets migrating through viscous disks (e.g. \citealt{Duffell(2014),DurmannKley(2015)}), we assume $\ap$ evolves in time according to
\be
\frac{\der \ap}{\der t} = -\frac{\ap}{t_{\rm mig}},
\label{eq:a_mig}
\ee
where
\be
t_{\rm mig} = \Lam_{\rm mig} \max \left( 1 , \frac{\Mp}{2\pi \bar \Sg(\ap) \ap^2} \right) \tv,
\label{eq:t_mig}
\ee
and $\Lam_{\rm mig} \sim 1$ is a factor parameterizing the uncertainty in $t_{\rm mig}$.  When $\Mp < 2 \pi \bar \Sg(\ap,t) \ap^2$, a smaller (larger) $\Lam_{\rm mig}$ value corresponds to a migration timescale $t_{\rm mig}$ shorter (longer) than the disk's viscous timescale $\tv$, parameterizing the effects seen in \cite{Duffell(2014)}.

For photo-ionized disks (Sec.~\ref{sec:Photo}), Eq.~\eqref{eq:a_mig} applies to planets in the outer disk ($r > \rc$).  We neglect the migration of planets in the inner depleted disk ($r < \rc$).

\subsection{Planet Interactions with Outer Disk}
\label{sec:PlanetOutDisk}

As discussed in Section~\ref{sec:Photo}, photoevaporation may deplete the inner disk ($r < \rc$) on a very short timescale.  If the planet's semi-major axis $\ap$ lies inside $\rc$, the mutual gravitational torques between the planet and the disk are modified.  As noted in Section~\ref{sec:Photo}, we neglect any misalignment between the inner ($r < \rc$) and outer ($r > \rc$) disks, since the timescale over which the inner disk is depleted is much shorter than the age of the system.  The precession rate of $\blp$ around $\ld$ due to the mass of the inner disk,
\be
\bompdl \simeq \frac{\pi \bar \Sg(\ap,t) \ap^2}{\Ms h_{\rm p}} \sqrt{ \frac{G \Ms}{\ap^3} },
\label{eq:bompdl}
\ee
is diminished due to the inner disk's rapid depletion from photoevaporation (see Sec.~\ref{sec:Photo}).  Instead, the precession of $\blp$ around $\ld$ is mainly governed by the torque on the planet from the outer disk ($r > \rc$):
\begin{align}
\bbT_{\rm pd>} &\simeq - \int_{\rc}^{\rout} \left( \frac{3 G \Mp \ap^2}{4 r^3} \right) (\blp \bcdot \ld)(\ld \btimes \blp) 2 \pi \Sg r \der r
\nonumber \\
&= - \Lp \bompdg (\blp \bcdot \ld)\ld \btimes \blp,
\label{eq:Tpdg}
\end{align}
where
\be
\bompdg \simeq \frac{3 \pi G \Sg(\rc,t)}{2(1+p) \rc} \sqrt{ \frac{\ap^3}{G \Ms} }
\label{eq:bompdg}
\ee
characterizes the precession frequency of the planet around the outer disk (assuming $\ap \ll \rc \ll \rout$).

The planet also exerts a back-reaction torque on the disk, causing the outer/inner disk to precess around the planet at the characteristic rates
\begin{align}
\bomdgp &\simeq \frac{3(5/2-p)}{4(1+p)} \left( \frac{\Mp}{\Ms} \right) \frac{\ap^2}{\rc^{1+p} \rout^{1-p}} \sqrt{ \frac{G \Ms}{\rout^3} },
\label{eq:bomdgp} \\
\bomdlp &\simeq \frac{5/2-p}{2 \Dg_{\rm p}/\ap} \left( \frac{\Sgc \rc^p \ap^{1-p}}{\Sgout \rout} \right) \left( \frac{\Mp}{\Ms} \right) \sqrt{ \frac{G \Ms}{\rout^3} }.
\label{eq:bomdlp}
\end{align}
[Compare the scaling of Eq.~\eqref{eq:bomdgp} to~\eqref{eq:bomds}, and \eqref{eq:bomdlp} to \eqref{eq:bomdp}]  The total planet-disk mutual precession rates are then
\begin{align}
\bompd &= \bompdg + \bompdl, 
\label{eq:bompd_photo} \\
\bomdp &= \bomdgp + \bomdlp.
\label{eq:bomdp_photo}
\end{align}

\subsection{Planet-Star and Planet-Binary Interactions}
\label{sec:PlanetStarBin}

\begin{figure}
\centering
\includegraphics[scale=0.7]{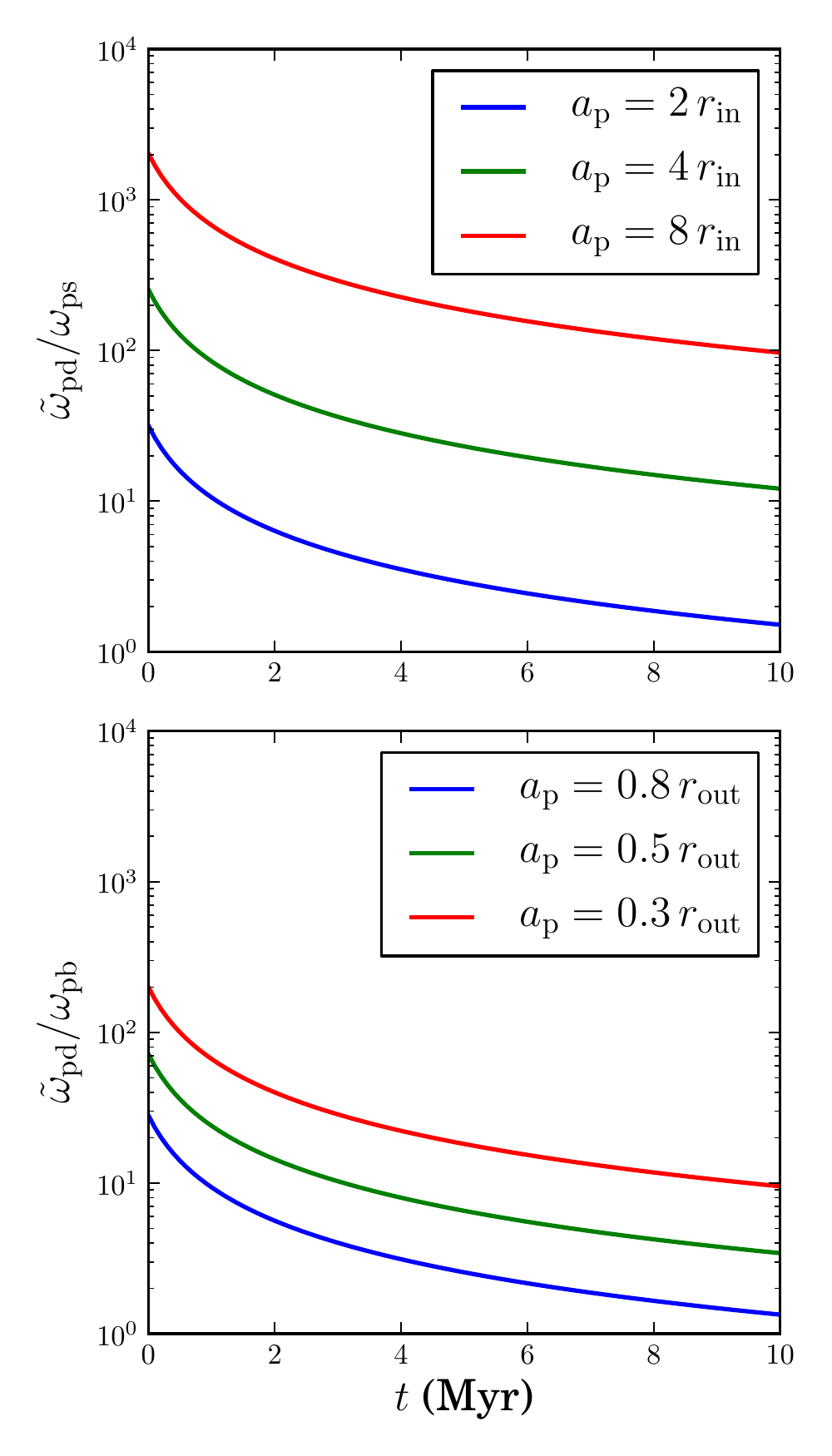}
\caption{
Ratio of the precession frequency of the planet driven by the disk $\bompd$ [Eq.~\eqref{eq:bompd}] to the precession frequency of the planet driven by the star $\omps$ [Eq.~\eqref{eq:omps}] and binary $\ompb$ [Eq.~\eqref{eq:ompb}] for different planetary semi-major axis $\ap$, with $p = 3/2$ and all other parameters canonical.  This plot shows the planet is tightly coupled to the disk, so we may approximate $\blp \simeq \ld$.  We assume $h_{\rm p} = 0.1$.
  }
\label{fig:bompd}
\end{figure}

The oblate central star exerts a torque on the planet, given by
\be
\bT_{\rm ps} = - \Lp \omps (\blp \bcdot \bs)\bs \btimes \blp,
\label{eq:Tps}
\ee
where
\be
\omps = \frac{3 \kq}{2} \bOmst^2 \left( \frac{\Rs}{\ap} \right)^2 \sqrt{ \frac{G \Ms}{\ap^3} }
\label{eq:omps}
\ee
characterizes the precession frequency of $\blp$ around $\bs$.  The back-reaction torque on the star from the planet causes $\bs$ to precess around $\blp$ at a characteristic rate
\begin{align}
\omsp &= (\Lp/S) \omps
\nonumber \\
&= \frac{3 \kq}{2 \ks} \bOmst \left( \frac{\Mp}{\Ms} \right) \frac{ \sqrt{ G \Ms \Rs^3 } }{\ap^3}.
\label{eq:omsp}
\end{align}

The binary companion also exerts torque on the planet:
\be
\bT_{\rm pb} = -\Lp \ompb (\blp \bcdot \blb)\blb \btimes \blp,
\ee
where
\be
\ompb = \frac{3 \Mb}{4 \Ms} \left( \frac{\ap}{\ab} \right)^3 \sqrt{ \frac{G \Ms}{\ap^3} }
\label{eq:ompb}
\ee
characterizes the precession frequency of $\blp$ around $\blb$.  Because the binary has orbital angular momentum $L_{\rm b} \gg \Lp$, the back reaction torque on the binary from the planet is neglected.

As discussed in Sections~\ref{sec:PlanetDiskNG}-\ref{sec:PlanetOutDisk}, the dominant planet-disk coupling involves mutual precession, with characteristic frequency $\bompd$.  For homologeously evolving disks, comparing $\bompd$ [Eq.~\eqref{eq:bompd}] to $\omps$ and $\ompb$, we see (assuming $p = 3/2$)
\begin{align}
\frac{\bompd}{\omps} = \ & 7.27 \left( \frac{\kq}{0.1} \right)^{-1} \left( \frac{h_{\rm p}}{0.1} \right)^{-1} \frac{\bMd}{\bMs}
\nonumber \\
&\times \left( \frac{\bOmst}{0.1} \right)^{-2} \frac{\brin^{5/2}}{\brout^{1/2} \bRs^2} \left( \frac{\ap}{\rin} \right)^{5/2},
\label{eq:pdps} \\
\frac{\bompd}{\ompb} = \ &7.20 \left( \frac{h_{\rm p}}{0.1} \right)^{-1} \frac{\bMd \bab^3}{\bMb \brout^3} \left( \frac{\rout}{\ap} \right)^{5/2},
\label{eq:pdpb}
\end{align}
where we have used Eqs.~\eqref{eq:p} and~\eqref{eq:Md} to relate $\bar \Sg(\ap,t)$ to $\Md$.  Figure~\ref{fig:bompd} plots the ratios~\eqref{eq:pdps} and~\eqref{eq:pdpb} for a standard disk model.  We see for most values of $\ap$ (with $\ap \gtrsim \text{a few} \ \rin$ and $\ap \ll \rout$), $\bompd \gg \omps, \ompb$ over the disk's lifetime.  This allows us to make the simplifying assumption $\blp(t) \simeq \ld(t)$ when $\Sg$ evolves homologously.  We note that in certain situations, a secular resonance between a planet, disk, and binary may greatly increase the planet-disk inclination \citep{LubowMartin(2016),Martin(2016)}, breaking the assumption that $\blp \simeq \ld$.

When the planet lies in the inner region ($\ap < \rc$) of a photo-ionized disk, we have (assuming $p = 3/2$)
\begin{align}
\frac{ \bompdg }{\omega_{\rm ps}} = & \ 2.1 \times 10^{-5} \left( \frac{\kq}{0.1} \right)^{-1} \left( \frac{\bOmst}{0.1} \right)^{-2} \left( \frac{\rout}{25 \, \rc} \right)^{5/2} 
\nonumber \\
&\times \frac{\brin^5 \bMd}{\bRs^2 \brout^3 \bMs} \left( \frac{\ap}{\rin} \right)^5,
\label{eq:PStoPDout} \\
\frac{ \bompdg }{\omega_{\rm pb}} = & \ 1.3 \times 10^3 \frac{\bMd \bab^3}{\bMb \brout^3} \left( \frac{\rout}{25 \, \rc} \right)^{5/2}.
\label{eq:PBtoPDout}
\end{align}
Clearly, the planet-outer disk precession frequency greatly exceeds the planet-binary precession frequency when $\ap < \rc$.  However, the ratio $\bompdg/\omps$ depends depends sensitively on the distance of the planet from the star.  The planet's full response to the star, disk and binary will need to be taken into account when the inner disk is depleted.

The star's response to the planet is important in the context of planet-star-disk-binary dynamics.  Comparing $\omsp$ [Eq.~\eqref{eq:omsp}] to $\bomsd$ [Eq.~\eqref{eq:bomsd}], we have (assuming $p = 3/2$)
\be
\frac{\omsp}{\bomsd} = 3.5 \left( \frac{\Mp}{1 \, \MJ} \right) \frac{\brout^{1/2}}{\bMd \brin^{1/2}} \left( \frac{\rin}{\ap} \right)^3.
\label{eq:spsd}
\ee
Equation~\eqref{eq:spsd} shows $\omsp \gtrsim \bomsd$ near the end of the disk's lifetime ($\bMd \ll 1$), when the planet lies close to the disk's inner truncation radius.  More interesting is the magnitude of $\omsp$ compared to $\bomdb$ [Eq.~\eqref{eq:bomdb}] (assuming $p = 3/2$):
\be
\frac{\omsp}{\bomdb} = 333 \left( \frac{2 \kq}{\ks} \right) \left( \frac{\bOmst}{0.1} \right) \left( \frac{\Mp}{1 \, \MJ} \right) \frac{\bab^3 \Rs^{3/2}}{\bMb \brin^3 \brout^{3/2}} \left( \frac{\rin}{\ap} \right)^3.
\label{eq:spdb}
\ee
Equation~\eqref{eq:spdb} shows for a substantial region of parameter space, $\omsp \gtrsim \bomdb$.  This implies that \textit{a close-in massive planet can suppress secular resonance}.  The next section explores different formation scenarios of close-in massive planets (hot Jupiters), and their implications to spin-orbit misalignments generated via star-disk-binary interactions.

\section{Inclination Evolution of Planet-Star-Disk-Binary Systems}
\label{sec:DynPSDB}

This section explores how the formation and migration of a gas giant in protoplanetary disks affects the generation of primordial spin-orbit misalignments through star-disk-binary interactions.  The core-accretion scenario assumes a gas giant forms following the run-away accretion of protoplanetary disk gas onto a $\sim 10 \, \ME$ core (e.g. \citealt{Pollack(1996)}).  After the formation of the massive planet, we consider three different models for its evolution through the disk.  The first assumes the planet forms in-situ, with the planet's semi-major axis $\ap$ fixed in time (Secs.~\ref{sec:EarlyInSitu}-\ref{sec:LateInSitu}), the second models the formation of a hot Jupiter via Type II migration (Sec.~\ref{sec:Mig}), while the last considers the system's dynamics after the hot Jupiter is left in a photo-ionized disk cavity (Sec.~\ref{sec:PlanetPhoto}).

We will frequently refer to the angles $\tg_{\rm ps}$, $\tg_{\rm pd}$, and $\tg_{\rm pb}$ throughout this section, defined as
\begin{align}
\tg_{\rm ps} &= \cos^{-1} (\blp \bcdot \bs),
\label{eq:tgps} \\
\tg_{\rm pd} &= \cos^{-1} (\blp \bcdot \ld),
\label{eq:tgpd} \\
\tg_{\rm pb} &= \cos^{-1}(\blp \bcdot \blb),
\label{eq:tgpb}
\end{align}
which define the mutual planet-star, planet-disk, and planet-binary inclinations, respectively.

\subsection{Early In-Situ Formation of Hot-Jupiters}
\label{sec:EarlyInSitu}

\begin{figure}
\centering
\includegraphics[scale=0.6]{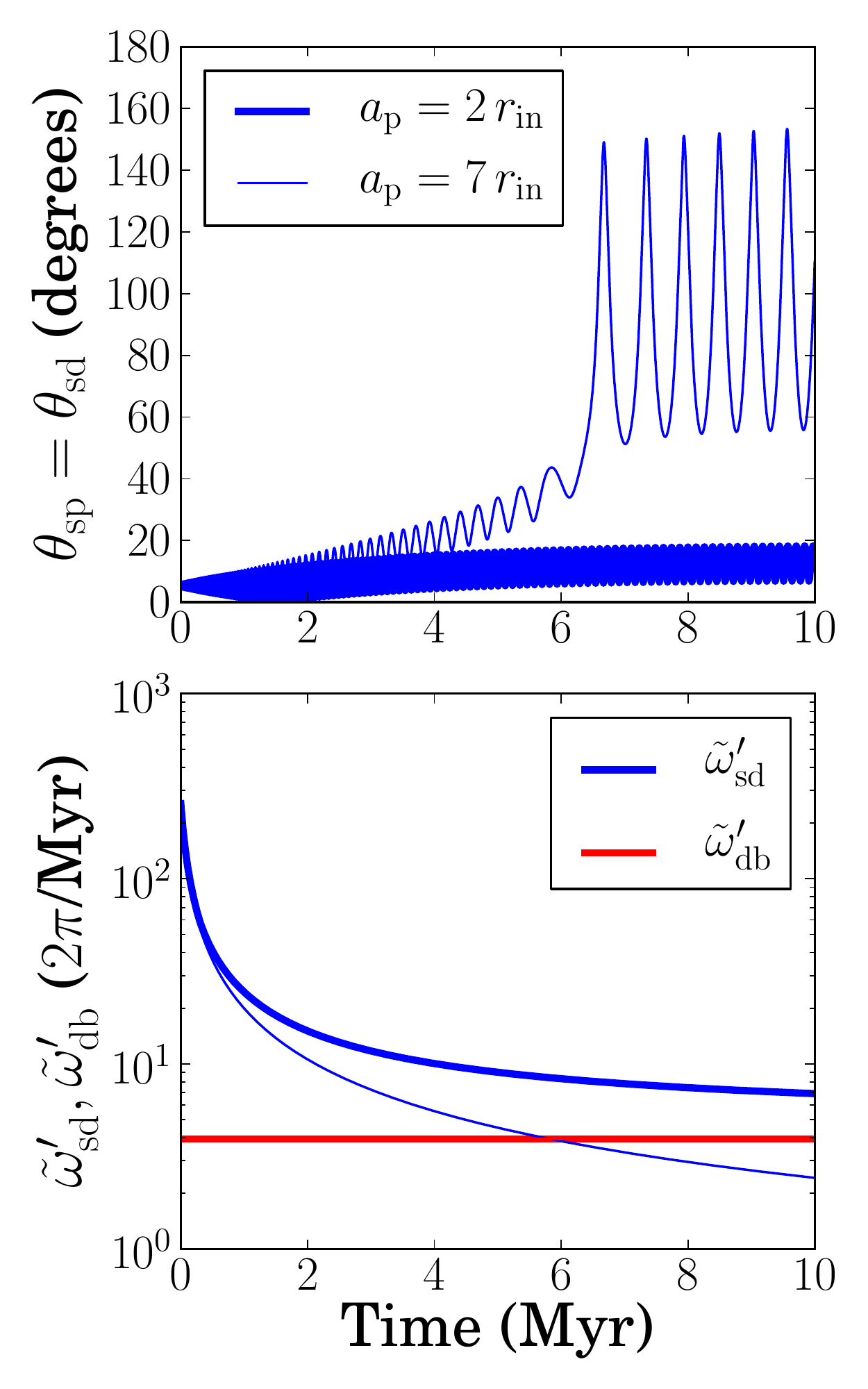}
\caption{
Evolution of the star-disk/star-planet angle $\theta_{\rm sd} = \theta_{\rm sp}$ (top panel), modified star-disk precession frequency
$\bomsd'$ [Eq.~\protect \eqref{eq:bomsd_p}] and modified disk-binary precession frequency $\bomdb'$ [Eq.~\protect \eqref{eq:bomdb_p}] (bottom panel) 
for planetary semi-major axis of $\ap = 2 \, \rin = 0.0736 \, \text{au}$ (thick lines) and $\ap = 7 \, \rin = 0.258 \, \text{au}$ (thin lines).  
We take all parameters to be canonical except $\bOmst = 0.03$, $p = 3/2$, $\tv = 0.1 \, \text{Myr}$, with $\tp = 0.3 \, \text{Myr}$ and $\xi = 10$.  
Here, the planet's mass $\Mp = 0.201 \, \MJ$ when $\ap = 2 \, \rin$ and $\Mp = 0.514 \, \MJ$ when $\ap = 7 \rin$.  We take the initial star-disk inclination $\theta_{\rm sd}(0) = 5^\circ$ and disk-binary inclination $\theta_{\rm db}(0) = 60^\circ$ for all integrations.  The hot Jupiter formed in-situ (thick lines) does not experience appreciable excitations of spin-orbit misalignment.
  }
\label{fig:EarlyInSitu}
\end{figure}

\cite{Batygin(2016)} proposed Hot-Jupiters form in-situ in their protoplanetary disks over timescales shorter than 1 million years.  They argued that a $~10 \, \ME$ core with $\ap \lesssim 0.1 \, \text{au}$ may undergo run-away accretion early in the disk's lifetime.  This scenario may explain why hot Jupiters do not have close, low-mass planetary companions \citep{Batygin(2016),SpaldingBatygin(2017)}.  Here we explore how in-situ formation affects the star-disk-binary system dynamics.  

For early formation of a gas giant, we assume the planet's mass is accreted from the disk within the planet's feeding zone.  Specifically, we take the time-dependent planetary mass to be
\be
\Mp(t) = \left\{ \begin{array}{cl}
2\pi \Sg(\ap,t) \ap \Dg \ap, & t < \tp \\
2\pi \Sg(\ap,\tp) \ap \Dg \ap & t \ge \tp
\end{array}
\right.
\label{eq:Mp_form}
\ee
where $\tp$ is the formation time,
\be
\Dg \ap = \xi \ap (\Mp/3 \Ms)^{1/3}
\label{eq:Dgap}
\ee
is the width of the planet's feeding zone, and $\xi$ is a free parameter.  This yields a final  ($t \ge \tp$) planetary mass of
\be
\frac{\Mp}{\Ms} = \frac{\xi^{3/2}}{3^{1/2}} \left[ \frac{2\pi \Sg(\ap, \tp) \ap^2}{\Ms} \right]^{3/2}.
\label{eq:Mp}
\ee
This model neglects accretion of gas onto the planet due to the viscous transport of disk material across the planet's gap.  This is a reasonable approximation, since simulations show the accretion rate onto a planet undergoing run-away gas accretion is typically much greater than the global accretion rate of the disk onto the host star (e.g. \citealt{PapaloizouNelson(2005),D'AngeloLubow(2008),AyliffeBate(2009),TanigawaTanaka(2016)}) 

Equation~\eqref{eq:Mp} with $p = 3/2$ gives a final planetary mass for a hot Jupiter formed in-situ of
\be
\frac{\Mp}{\Ms} = \frac{1.93 \times 10^{-3}}{(1+\tp/\tv)^{3/2}} \left( \frac{\xi}{10} \right)^{3/2} \frac{\bar M_{\rm d0}^{3/2}}{\bar \Ms^{3/2} \brout^{3/4}} \left( \frac{\ap}{0.1 \, \text{au}} \right)^{3/4}.
\label{eq:Mp_EarlyInSitu}
\ee
Even with a large feeding zone $(\xi = 10)$, we see that the hot Jupiter must form at a time $\tp \lesssim \text{few} \times \tv \sim 1 \, \text{Myr}$ to attain mass $\Mp \sim 1\, \MJ$.

We assume in this subsection that $\blp(t) \simeq \ld(t)$, since $\bompd \gg \omps, \ompb$ when the planet is embedded in the disk (see Fig.~\ref{fig:bompd}).  The evolution equations for $\bs$ and $\ld = \blp$ become
\begin{align}
\frac{\der \bs}{\der t} &= - \bomsd' (\bs \bcdot \ld) \ld \btimes \bs, 
\label{eq:dsdt_full} \\
\frac{\der \ld}{\der t} &= - \bomds' (\ld \bcdot \bs) \bs \btimes \ld - \bomdb' (\ld \bcdot \blb) \blb \btimes \ld,
\label{eq:dlddt_full}
\end{align}
where
\begin{align}
\bomsd' &= \bomsd + \omsp,
\label{eq:bomsd_p} \\
\bomds' &= \bomds + (\Lp/\Ld)\omps,
\label{eq:bomds_p} \\
\bomdb' &= \bomdb + (\Lp/\Ld) \ompb,
\label{eq:bomdb_p}
\end{align}
are the mutual star-disk-binary precession frequencies modified by the presence of a massive planet.  

Figure~\ref{fig:EarlyInSitu} shows the evolution of $\theta_{\rm sd} = \tg_{\rm sp}$ (top panel) and precession frequencies $\bomsd'$ and $\bomdb'$ (bottom panel).  We see when the planet forms too close to its host star ($\ap = 2 \, \rin$), $\bomsd'$ is always larger than $\bomdb'$ thoughout the disk evolution, the system averts secular resonance, and no significant spin-orbit misalignment is generated.  In other words, a close-in giant planet makes $\bs$ closely follow $\ld \simeq \blp$.  When the planet forms further from it's host star ($\ap = 7 \, \rin$), $\omsp$ is reduced, and the system goes through secular resonance, and significant $\tg_{\rm sp}$ is achieved.  The star-planet-disk-binary system may undergo secular resonance when the planet forms at a sufficiently large $\ap$, so that $\omsp \lesssim \bomdb$.  This inequality gives a lower bound for $\ap$:
\be
\ap \gtrsim \left[ \frac{2(4-p)\kq}{(5/2-p)\ks} \right]^{1/3} \bOmst^{1/3} \left( \frac{\Mp}{\Mb} \right)^{1/3} \left( \frac{\Rs}{\rout} \right)^{1/2} \ab.
\label{eq:ap_lower_gen}
\ee
Taking $p = 3/2$, we have
\be
\ap \gtrsim 0.23 \left( \frac{2 \kq}{\ks} \right)^{1/3} \left( \frac{\bOmst}{0.1} \right)^{1/3} \left( \frac{\Mp}{1 \, \MJ} \right)^{1/3} \frac{\bRs^{1/2} \bab}{\bMb^{1/3} \brout^{1/2}} \, \text{au}.
\label{eq:ap_lower_3/2}
\ee
 Thus, only giant planets formed at large distances ($\ap \gtrsim 0.2 \, \text{au}$) have any chance of experiencing excitation of spin-orbit misalignment from star-disk-binary interactions.  Lower mass planets ($\Mp \lesssim 0.1 \, \MJ$) formed around slowly-spinning stars ($\bOmst \lesssim 0.03$) with close binary companions ($\ab \lesssim 200 \, \text{au}$) may experience excitation of spin-orbit misalignments when $\ap \gtrsim 0.04 \, \text{au}$, but this is a very limited region of the parameter space of observed star-disk-binary systems.  We conclude significant $\tg_{\rm sp}$ is unlikely to be excited when a HJ forms in-situ early ($\tp \lesssim 1 \, \text{Myr}$).

\subsection{Late In-Situ Formation of Hot-Jupiters}
\label{sec:LateInSitu} 

\begin{figure*}
\centering
\includegraphics[scale=0.6]{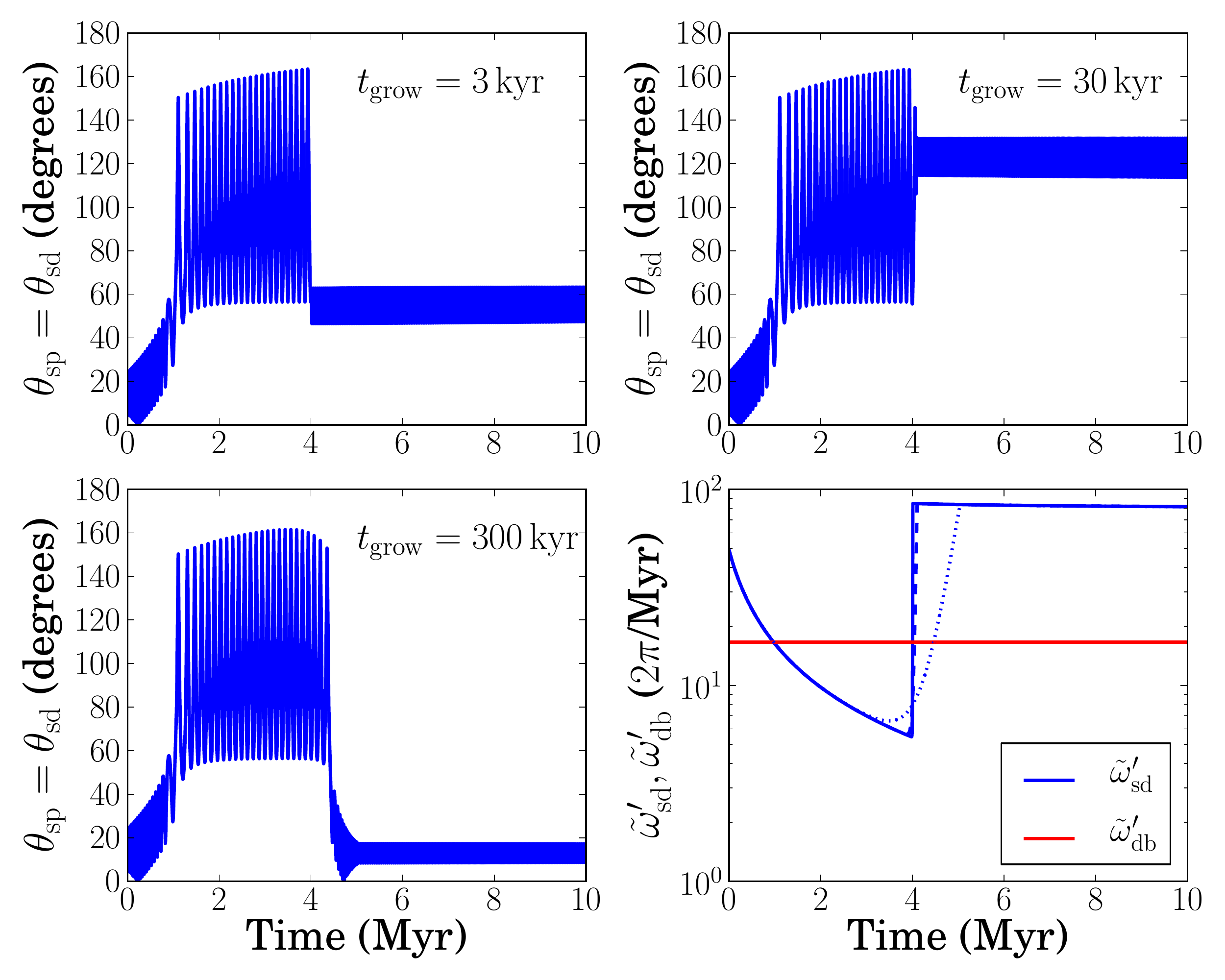}
\caption{ Evolution of the star-planet/star-disk inclination $\tg_{\rm sp} = \tg_{\rm sd}$ in the late in-situ model, for different planet mass growth timescale $t_{\rm grow}$ [Eq.~\eqref{eq:t_grow}] as indicated (top and bottom left panels).  The bottom right panel shows the modified star-disk precession frequency
$\bomsd'$ [Eq.~\protect \eqref{eq:bomsd_p}] and modified disk-binary precession frequency $\bomdb'$ [Eq.~\protect \eqref{eq:bomdb_p}] for $t_{\rm grow} = 3 \, \text{kyr}$ (solid), $t_{\rm grow} = 30 \, \text{kyr}$ (dashed), and $t_{\rm grow} = 300 \, \text{kyr}$ (dotted).  All parameter values are canonical [Eq.~\eqref{eq:pars}] except the binary's semi-major axis $\ab = 200 \, \text{au}$.  The planet has a semi-major axis of $\ap = 2 \, \rin = 0.0736 \, \text{au}$, and forms at time $\tp = 4 \, \Myr$.  The adiabatic parameter $A$ [Eq.~\eqref{eq:A}] takes values of $A = 0.312$ ($t_{\rm grow} = 3 \, \text{kyr}$), $A = 3.12$ ($t_{\rm grow} = 30 \, \text{kyr}$), and $A = 31.2$ ($t_{\rm grow} = 300 \, \text{kyr}$).  Large star-planet/disk inclinations are maintained only when $A \lesssim \text{a few}$.
}
\label{fig:LateInSitu}
\end{figure*}

\cite{Boley(2016)} proposed that a $\sim 10 \, \ME$ core may form at orbital periods $\lesssim 10 \, \text{days}$ after a phase of dynamical instability in a short-period ($\lesssim 200 \, \text{days}$) multi-planet system.  If this critical core forms late in the disk's lifetime ($\tp \gtrsim 1 \, \Myr$), the planet cannot accrete much of the disk's mass locally [Eq.~\eqref{eq:Mp_EarlyInSitu}].  Therefore, we assume a hot Jupiter formed late ($\tp \gtrsim 1 \, \Myr$) in the disk's lifetime grows primarily from disk mass advected through the planets gap.  Simulations show the accretion rate of viscously advected disk-mass onto a gap-opening planet may be written as
\be
\frac{\der \Mp}{\der t} = -\eta \frac{\der \Md}{\der t},
\label{eq:dMpdt_visc}
\ee
where $\eta \sim 0.7-0.9$ depending on the planet's mass and local disk properties (e.g. \citealt{LubowD'Angelo(2006)}).  The accretion rate~\eqref{eq:dMpdt_visc} will cause the planet's mass $\Mp$ to grow on a timescale
\begin{align}
&t_{\rm grow} \equiv \frac{\Mp}{\der \Mp/\der t} \sim \frac{\tv}{\eta} \left( \frac{\Mp}{\Md} \right)
\nonumber \\
&= 2.15 \, \text{kyr} \left( \frac{\tv}{0.5 \, \Myr} \right) \left(\frac{0.7}{\eta} \right) \left( \frac{\Mp}{10 \, \ME} \right) \bMd^{-1}.
\label{eq:t_grow}
\end{align}
Because in this model the planet's mass is accreted globally from the disk, we assume that $\Mp$ remains independent of the local disk properties (most notably the disk surface density near $r = \ap$), and prescribe $\Mp = \Mp(t)$ as
\be
\Mp(t) = \min \left[ 10 \, \ME \exp \left( \frac{t-\tp}{t_{\rm grow}} \right), 1 \, \MJ \right].
\label{eq:Mp_visc}
\ee
Notice our early in-situ formation model for hot Jupiters [Eq.~\eqref{eq:Mp}] fixes $\omsp = \text{constant}$ when $t \ge \tp$, while our late in-situ formation model [Eq.~\eqref{eq:Mp_visc}] causes $\omsp$ to grow until $\Mp = 1 \, \MJ$.

Because the late in-situ formation of a hot Jupiter causes an increase of $\bomsd'$ after formation, we expect the system to encounter a second secular resonance (when $\omsp \sim \bomdb$) if the system undergoes an initial secular resonance (when $\bomsd \sim \bomdb$) before the planet forms.  The timescale of this second resonance crossing is of order $t_{\rm grow}$.  If $t_{\rm grow}$ is sufficiently long compared to $(\bomdb)^{-1}$, a large amount of angular momentum may be exchanged throughout the planet-star-disk-binary system during the resonance crossing, significantly influencing the final star-planet/disk inclinations $\tg_{\rm sp} = \tg_{\rm sd}$.  If $t_{\rm grow}$ is comparable or shorter than $(\bomdb)^{-1}$, the system cannot exchange much angular momentum during its period of secular resonance, effectively freezing the star-planet/disk inclination at the time the planet forms ($\tg_{\rm sp}(t) \approx \tg_{\rm sp}(\tp)$ when $t \ge \tp$).  We introduce the adiabaticity parameter
\be
A = t_{\rm grow} \bomdb.
\label{eq:A}
\ee
When $A \gg 1$, we expect a large amount of angular momentum to be exchanged between the stellar spin and the planet/disk orbital angular momenta.

Figure~\ref{fig:LateInSitu} shows the evolution of $\tg_{\rm sp} = \tg_{\rm sd}$ (top and bottom left panels) and precession frequencies $\bomsd'$ and $\bomdb'$ (bottom right panel) using our late in-situ hot Jupiter formation model, with $t_{\rm grow}$ indicated.  In the top two panels, $A \lesssim \text{a few}$, so the system's second secular resonance does not allow a significant amount of angular momentum to be transferred from the stellar spin to the planet/disk angular momenta.  As a result, the star-planet/disk inclinations freeze to $\tg_{\rm sp} \approx 60^\circ$ (top left) and $\tg_{\rm sp} \approx 120^\circ$ (top right) after the planet forms ($\tp = 4 \, \Myr$).  In the bottom left panel, $A = 31.2 \gg 1$, so the star-planet/disk inclination settles down to $\tg_{\rm sp} \approx 10^\circ$ after the second resonance crossing.

\subsection{Formation of Hot-Jupiters through Type-II Migration}
\label{sec:Mig}

\begin{figure}
\centering
\includegraphics[scale=0.5]{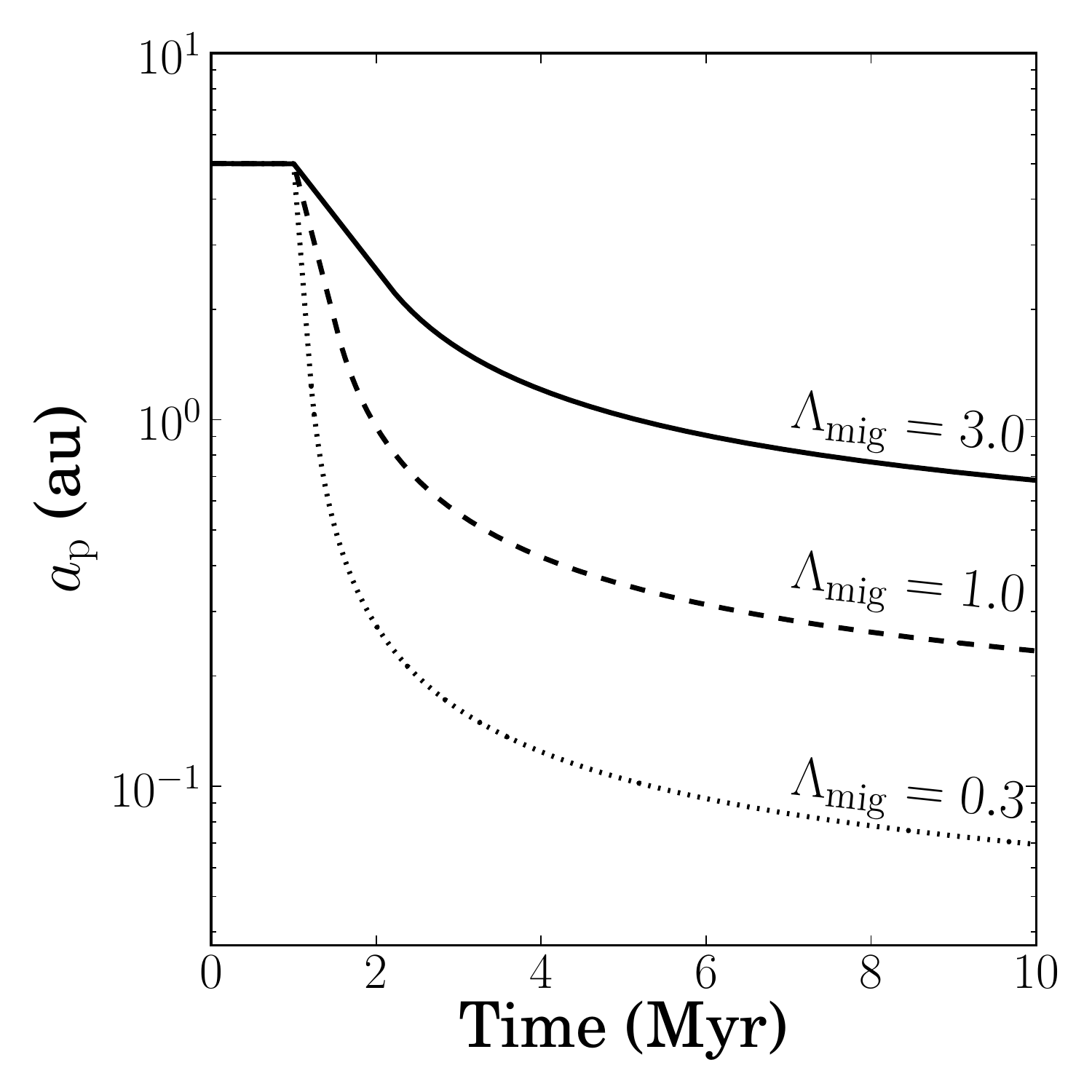}
\caption{
Evolution of the planet's semi-major axis $\ap$ with time, for different $\Lam_{\rm mig}$ values as  indicated [see Eq.~\eqref{eq:t_mig}].  We assume the planet forms at $\tp = 1 \, \text{Myr}$ and $\ap = 5 \, \text{au}$ with $\Mp = 0.93 \, \MJ$, assuming $\xi = 4$ (assuming $p = 1$, and canonical disk parameters).  The planet migrates to $\ap = 0.07 \, \text{au}$ ($\Lam_{\rm mig} = 0.3$, fast migration), $\ap = 0.23 \, \text{au}$ ($\Lam_{\rm mig} = 1.0$, moderate migration), and $\ap = 0.68 \, \text{au}$ ($\Lam_{\rm mig} = 3.0$, slow migration).
  }
\label{fig:ap_mig}
\end{figure}

\begin{figure*}
\includegraphics[scale=0.5]{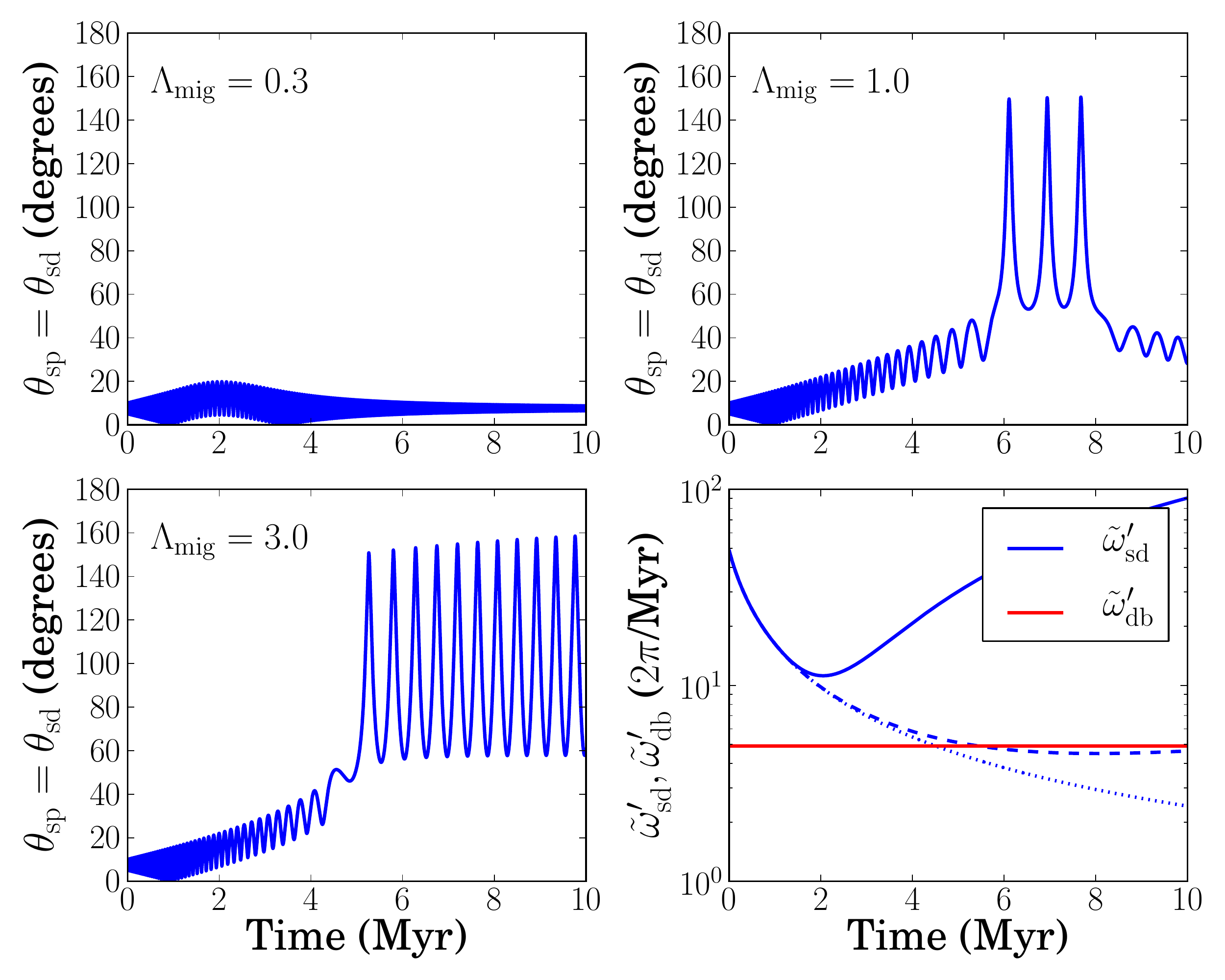}
\caption{
Evolution of the star-disk/star-planet angle $\theta_{\rm sd} = \theta_{\rm sp}$ (top and bottom left panels) and precession frequencies 
$\bomsd'$ [Eq.~\protect \eqref{eq:bomsd_p}] and $\bomdb'$ [Eq.~\protect \eqref{eq:bomdb_p}] (bottom right panel)  with time,
for the $\Lam_{\rm mig}$ values indicated.  In the bottom right panel, the different lines correspond to fast migration ($\Lam_{\rm mig} = 0.3$, dotted), moderate migration ($\Lam_{\rm mig} = 1.0$, dashed), and slow migration ($\Lam_{\rm mig} = 3.0$, solid).  We take all parameters to be cannonical with $p = 1$ and $\tv = 0.5 \, \text{Myr}$.  The planet forms at $\tp = 1 \, \text{Myr}$ and $\ap = 5 \, \text{au}$ with $M_{\rm p} = 0.931 \, \MJ$ [assuming $\xi = 4$ in Eq.~\eqref{eq:Mp}].  We take $\theta_{\rm sd}(0) = 5^\circ$ and $\theta_{\rm db}(0) = 60^\circ$ in all integrations.  See Fig.~\ref{fig:ap_mig} for the $\ap$ evolution.  No appreciable $\tg_{\rm sp}$ is generated when a hot Jupiter is produced (see the $\Lam_{\rm mig} = 0.3$ case).
  }
\label{fig:Mig1}
\end{figure*}

\begin{figure}
\centering
\includegraphics[scale=0.6]{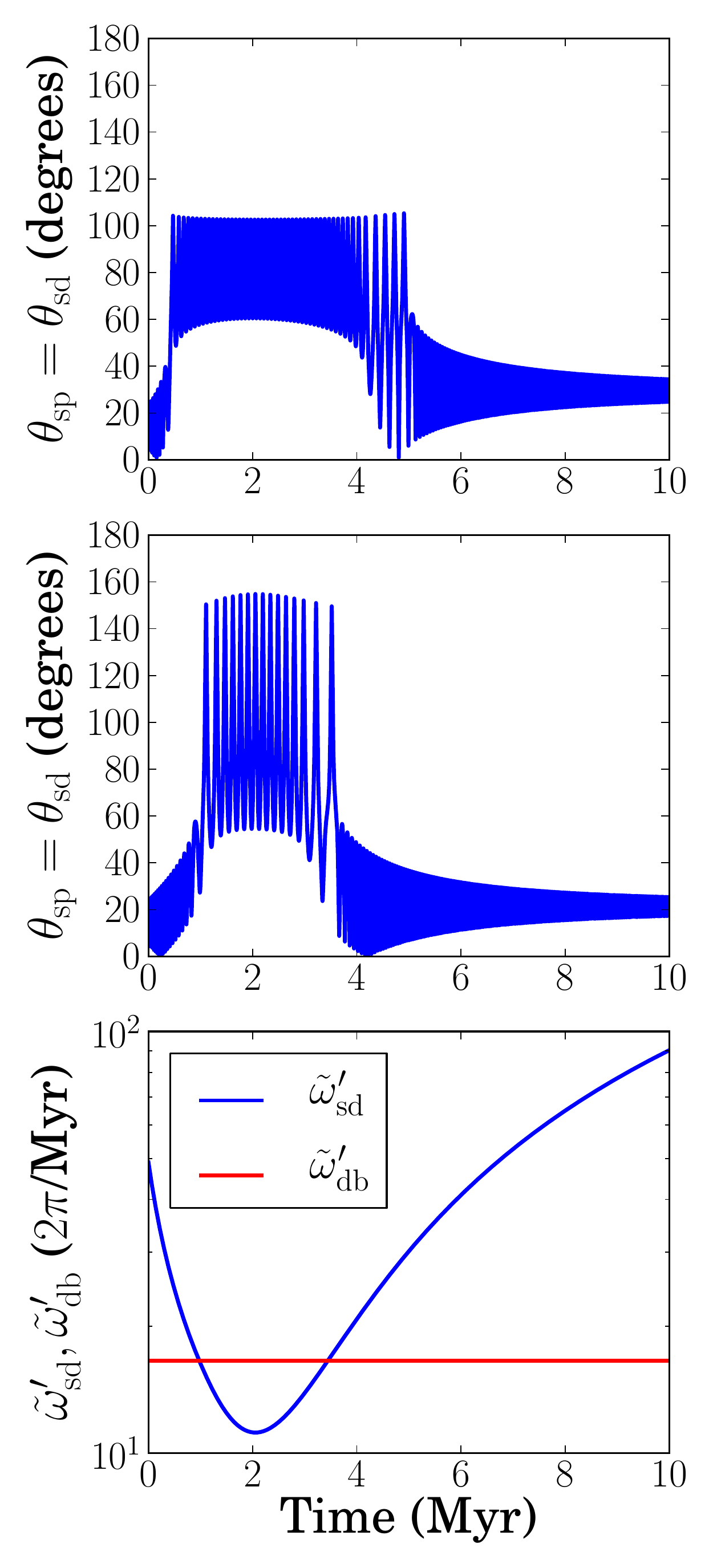}
\caption{
Evolution of the star-disk/star-planet angle $\theta_{\rm sd} = \theta_{\rm sp}$ (top and middle panels) and precession frequencies 
$\bomsd'$ [Eq.~\protect \eqref{eq:bomsd_p}] and $\bomdb'$ [Eq.~\protect \eqref{eq:bomdb_p}] (bottom panel)  with $\theta_{\rm db}(t)$ approximately equal to $\theta_{\rm db}(0) = 20^\circ$ (top) and $\theta_{\rm db}(0) = 60^\circ$ (middle), for $\Lam_{\rm mig} = 0.3$ and $\theta_{\rm sd}(0) = 5^\circ$ in all integrations.  We take all parameters to be canonical except $\ab = 200 \, \text{au}$ with $p = 1$ and $\tv = 0.5 \, \text{Myr}$.  The planet forms at $\tp = 1 \, \text{Myr}$ and $\ap = 5 \, \text{au}$ with $M_{\rm p} = 0.931 \, \MJ$ [assuming $\xi = 4$ in Eq.~\eqref{eq:Mp}].  See Fig.~\ref{fig:ap_mig} for the $\ap$ evolution.  Significant ($\tg_{\rm sp} \gtrsim 30$) spin-orbit misalignments are not sustained when a Jovian planet migrates close to it's host star after the star-disk-binary system experiences secular resonance ($\bomsd \sim \bomdb$).
  }
\label{fig:Mig2}
\end{figure}

We now consider the scenario where the giant planet forms at a large semi-major axis and subsequently undergoes Type-II migration.  The planet forms with a mass $\Mp$ given by Eq.~\eqref{eq:Mp_form}, where $\ap$ is fixed when $t \le \tp$, afterwards it migrates inwards according to Eqs.~\eqref{eq:a_mig} and~\eqref{eq:t_mig}.  For $p = 1$, Eq.~\eqref{eq:Mp} gives
\be
\frac{\Mp}{\Ms} = \frac{1.63 \times 10^{-3}}{(1+\tp/\tv)^{3/2}} \left( \frac{\xi}{4} \right)^{3/2} \frac{\bar M_{\rm d0}^{3/2}}{\bar M_\star^{3/2} \bar r_{\rm out}^{3/2}} \left( \frac{\ap(\tp)}{5 \, \text{au}} \right)^{3/2}.
\ee
Figure~\ref{fig:ap_mig} shows the semi-major axis evolution of a planet formed at $\tp = 1 \, \text{Myr}$ with $\ap(\tp) = 5 \, \text{au}$.  The shortest migration time parameter ($\Lam_{\rm mig} = 0.3$) leads to a hot Jupiter at the end of the disk's lifetime ($\ap \lesssim 0.1 \, \text{au}$ at $t = 10 \, \text{Myr}$), while longer migration time parameters allow $\ap$ to decrease and stop at a value $\gtrsim 0.1 \, \text{au}$.  Fig.~\ref{fig:ap_mig} also shows that most of the planet's migration occurs when the disk is young ($t \lesssim \text{few} \times \tv$), since the reduction of disk mass lengthens $t_{\rm mig}$ significantly [see Eq.~\eqref{eq:t_mig}]. 

Figure~\ref{fig:Mig1} plots $\tg_{\rm sp} = \theta_{\rm sd}$ (top and bottom left panels), $\bomsd'$, and $\bomdb'$ (bottom right panel) with time, for the $\Lam_{\rm mig}$ values indicated.  When $\Lam_{\rm mig} = 0.3$,  the planet quickly migrates close to its host star within the first few Myr's (see Fig.~\ref{fig:ap_mig}).  This causes $\bomsd'$ to increase in time after a few Myr's, ensuring that the secular resonance is never achieved (bottom right panel of Fig.~\ref{fig:Mig1}).  As a result, $\theta_{\rm sp}$ is not excited by star-disk-binary interactions (top left panel of Fig.~\ref{fig:Mig1}), and the hot Jupiter forms without spin-orbit misalignment.  When $\Lam_{\rm mig} = 1.0$, the planet migrates to an $\ap$ value so that $\bomsd' \sim \bomdb'$ after a few Myr's.  The star-disk-binary system proceeds to pass into and out of secular resonance, generating a planet with spin-orbit misalignment (top right panel of Fig.~\ref{fig:Mig1}).  However, this planet has become a warm Jupiter, with a final semi-major axis $\ap = 0.234 \, \text{au}$ at $t = 10 \, \text{Myr}$ (see Fig.~\ref{fig:ap_mig}).  When $\Lam_{\rm mig} = 3.0$, the planet stays sufficiently far from its host star so that secular resonance may occur without modification by $\omsp$, allowing $\theta_{\rm sp}$ to be excited by star-disk-binary interactions (bottom left panel of Fig.~\ref{fig:Mig1}).  The $\Lam_{\rm mig} = 3.0$ planet ends at a semi-major axis of $\ap = 0.683 \, \text{au}$, far too large to be considered a hot Jupiter.  

Figure~\ref{fig:Mig2} shows another example of the evolution of star-disk-binary system, in which a hot Jupiter forms via Type-II migration after secular resonance (when $\bomsd' \sim \bomdb'$).  A large $\tg_{\rm sp}$ is achieved while $\bomsd' \lesssim \bomdb'$.  However, once the planet migrates close enough to its host star so that $\omsp \gtrsim \bomdb$, the system passes through secular resonance again, and $\bs$ switches from precessing around $\blb$ to precessing around $\ld \simeq \blp$.  Although $\tgsd$ evolves to values significantly larger than the initial $\tgsd(0) = 5^\circ$, the final stellar obliquity is modest, and border on being considered coplanar ($\theta_{\rm sp} \lesssim 20^\circ$).  We see that even when the star-disk-planet-binary system does undergo secular resonance, the star-planet interaction significantly reduces spin-orbit misalignment after the planet has migrated near the vicinity of the host star.

\subsection{Hot Jupiters left in disk cavity from photoevaporation}
\label{sec:PlanetPhoto}

\begin{figure*}
\includegraphics[scale=0.5]{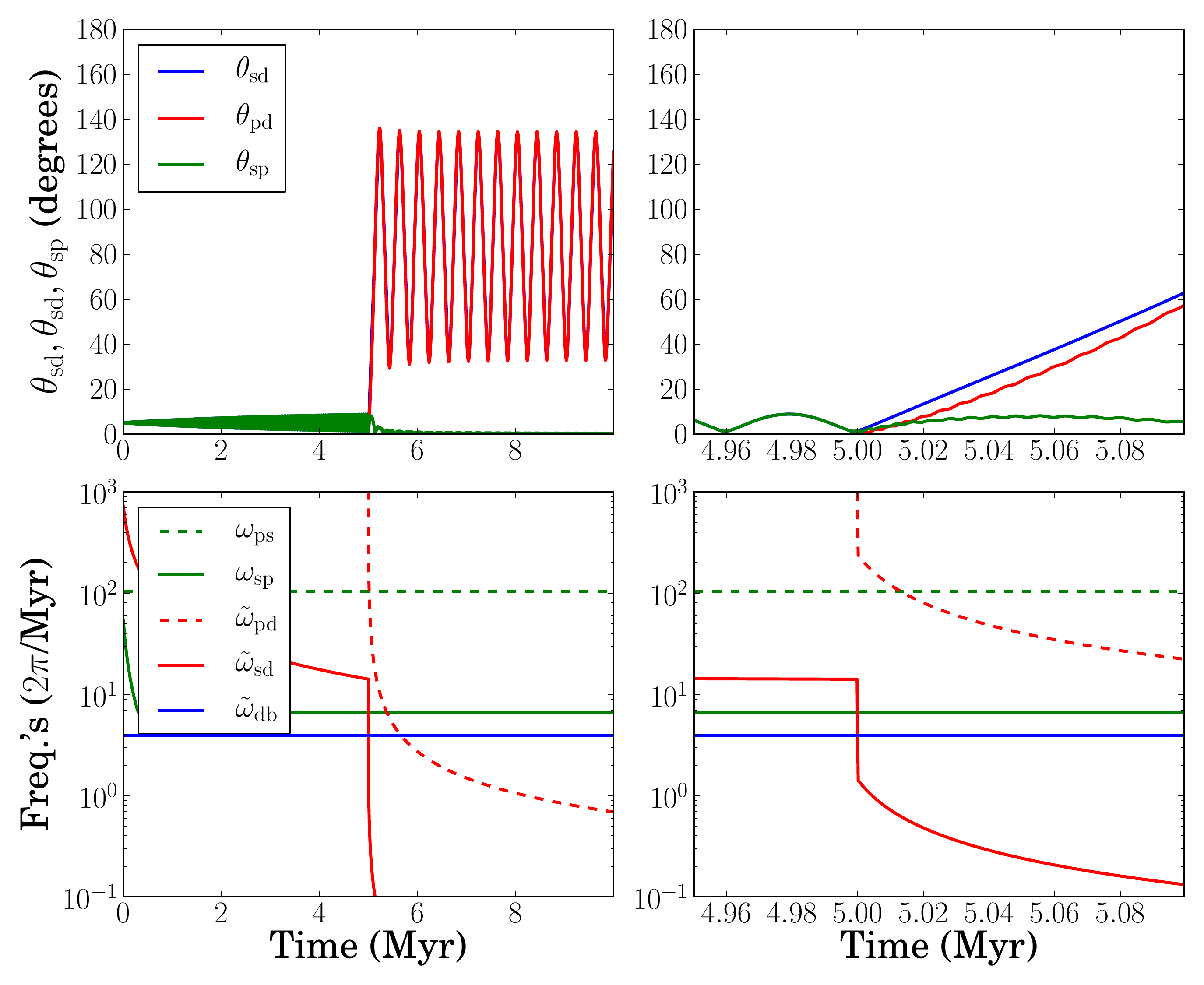}
\caption{
Star-disk-planet-binary evolution under a prescribed $\Sg$ depleation due to photoevaporation (see Sec.~\ref{sec:Photo}).  Top panels display $\theta_{\rm sd}$, $\theta_{\rm pd}$, and $\theta_{\rm ps}$ with time, bottom panels display numerous characteristic precession frequencies with time.  Left panels show the system's entire dynamical evolution over 10 Myr, right panels show the system's dynamics near $t \approx \tw = 5 \, \text{Myr}$.  All parameters are cannonical except $p = 3/2$ and $\tv = 0.1 \, \text{Myr}$, with $\tw = 5 \, \text{Myr}$, $t_{\rm v,in} = 0.01 \, \text{Myr}$, $\rc = 2 \, \text{au}$, $h_{\rm p} = 0.2$, $\ap(t) = \ap(0) = 4 \, \rin = 0.147 \, \text{au}$, $\tp = 0.3 \, \text{Myr}$, and $\Mp = 0.338 \, \MJ$ (assuming $\xi = 10$).  We take $\theta_{\rm sd}(0) = 5^\circ$ and $\theta_{\rm db}(t) \approx \theta_{\rm db}(0) = 60^\circ$ for all integrations.  Because $\bomsd' \gtrsim \bomdb'$ at $t \approx \tw$, the star-planet inclination $\tg_{\rm sp}$ stays negligible.
  }
\label{fig:photo_damp}
\end{figure*}

\begin{figure*}
\includegraphics[scale=0.5]{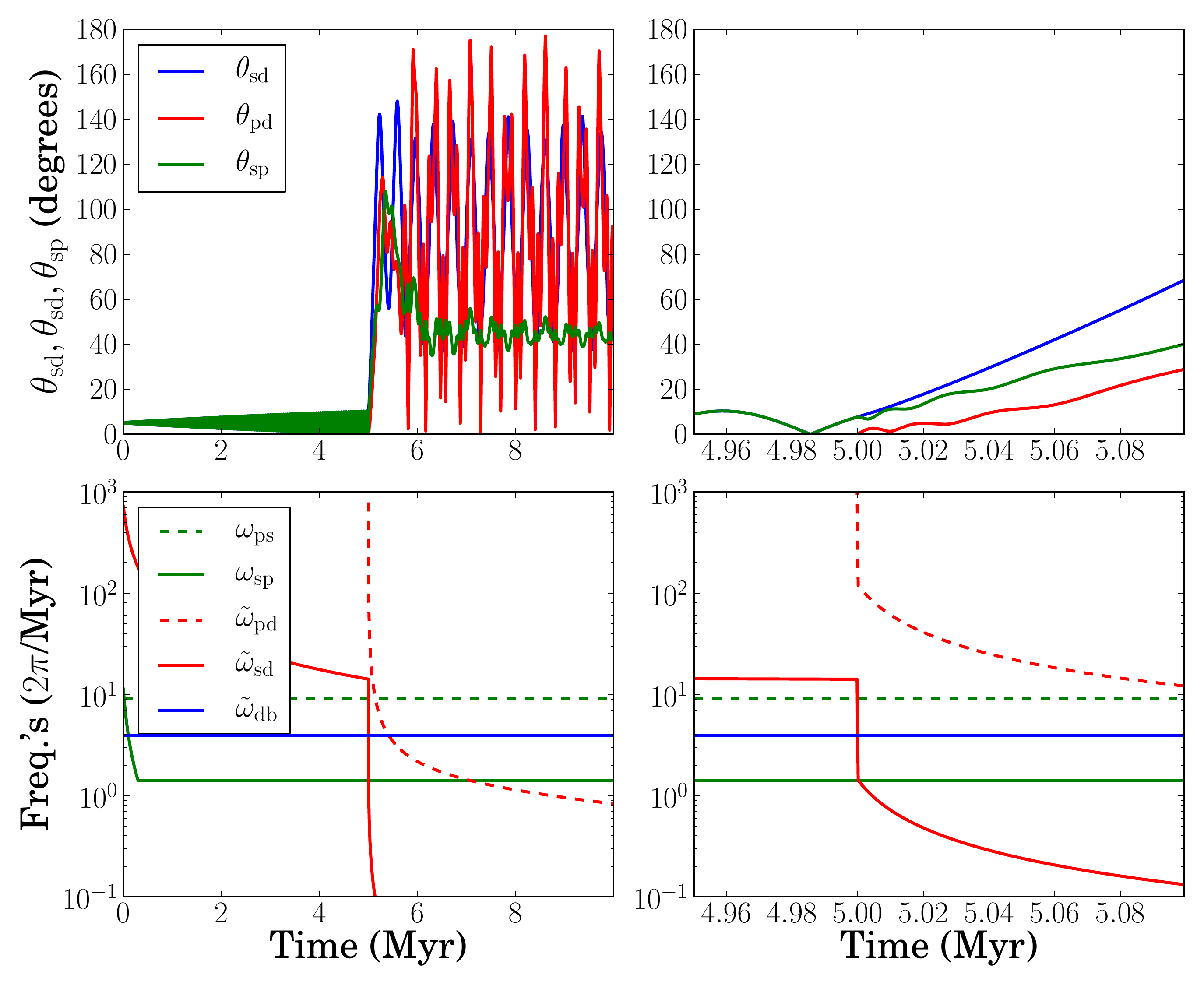}
\caption{
Same as Fig.~\protect\ref{fig:photo_damp} except $\ap(t) = \ap(0) = 8 \, \rin = 0.294 \, \text{au}$, $\tp = 0.3 \, \text{Myr}$ with $\Mp = 0.568 \MJ$ (assuming $\xi = 10$).  Because $\bomsd' \lesssim \bomdb'$ at $t \approx \tw$, the star-planet inclination $\tg_{\rm sp}$ is excited after $\tw$.
  }
\label{fig:photo_excite}
\end{figure*}

This section examines the fate of hot Jupiters in star-disk-binary systems when the disk's inner cavity is rapidly cleared by photoevaporation.  We adopt the $\Sg(r,t)$ prescription of Section~\ref{sec:Photo}.  For simplicity, we assume the hot Jupiters form in-situ, although they could have undergone Type-II migration before the inner disk is depleted at $t \approx \tw$.  Note that when the planet lies in the inner disk depleted by photoevaporation [see Eq.~\eqref{eq:t_mig}], radial migration is negligible, halting the hot Jupiter at $\ap \approx \ap(\tw)$.  We assume the planet forms at $t = \tp < \tw$ with mass $\Mp$ given by Eq.~\eqref{eq:Mp}.

The evolution of the planet-star-disk-binary system proceeds in two stages.  For $t \le \tw$, the planet is embedded in the ``full" disk, so $\blp \simeq \ld$ and Eqs.~\eqref{eq:dsdt_full}-\eqref{eq:dlddt_full} apply.  For $t > \tw$, the planet resides in a depleted disk cavity and $\bompd \lesssim \omps, \ompb$ [see Eqs.~\eqref{eq:PStoPDout}-\eqref{eq:PBtoPDout}], so $\blp$ and $\ld$ decouple and evolve separately.  The evolution equations for $t > \tw$ are
\begin{align}
\frac{\der \bs}{\der t} = & -\bomsd(\bs \bcdot \ld) \ld \btimes \bs - \omsp(\bs \bcdot \blp) \blp \btimes \bs, 
\label{eq:dsdt_photo} \\
\frac{\der \ld}{\der t} = & -\bomds (\ld \bcdot \bs) \bs \btimes \ld - \bomdp (\ld \bcdot \blp)\blp \btimes \ld 
\nonumber \\
&- \bomdb (\ld \bcdot \blb)\blb \btimes \ld,
\label{eq:dlddt_photo} \\
\frac{\der \blp}{\der t} = & -\bompd (\blp \bcdot \ld) \ld \btimes \blp - \omps (\blp \bcdot \bs) \bs \btimes \blp
\nonumber \\
&- \ompb (\blp \bcdot \blb)\blb \btimes \blp,
\label{eq:dlpdt_photo}
\end{align}
where $\bompd$ and $\bomdp$ are given by Eqs.~\eqref{eq:bompd_photo} and~\eqref{eq:bomdp_photo}.

Before $\tw$, the disk/planet is strongly coupled to the star ($\bomsd' \gg \bomdb'$).  Immediately  following $\tw$, the inner disk's rapid depletion causes $\bomsd$ to fall well below $\omsp$.  The main coupling allowing $\bs$ to track $\blp$ and $\ld$ is $\omsp$, while the main external forcing trying to disrupt the mutual planet-star-disk coupling is $\bomdb$.  Since $\bompd$ is typically much larger than the other frequencies during this time, $\blp$ and $\ld$ are coupled.  Whether $\bs$ is allowed to become significantly misaligned with $\blp$ and $\ld$ after $\tw$ depends on the magnitude of $\omsp$ compared to $\bomdb$:
\begin{enumerate}
\item If $\omsp \gtrsim \bomdb$, the planet star-disk coupling is stronger than the disk-binary coupling working to misalign the planet-star-disk system.  The stellar spin $\bs$ stays aligned $\blp$ and $\ld$, and stellar obliquity is not excited at $t \approx \tw$.
\item If $\omsp \lesssim \bomdb$, the planet star-disk coupling is weaker than the disk-binary coupling.  The stellar spin $\bs$ decouples from $\blp$ and $\ld$, and stellar obliquities are excited at $t \approx \tw$.
\end{enumerate}

Soon after $\tw$, $\bompd \approx \bompdg$ falls well below $\omps$, and $\blp$ decouples from $\ld$ but remains strongly coupled $\bs$.  The stellar obliquities excited in planet-star-disk-binary systems over the disk's lifetime depends on the magnitude of $\omsp$ compared to $\bomdb$.

Figure~\ref{fig:photo_damp} displays the evolution of the star-disk-planet-binary system for $\ap = 4 \, \rin = 0.147 \,  \text{au}$, which implies $\omsp \gtrsim \bomdb$.  We assume $\Sg(r,t)$ evolves under prescription~\eqref{eq:Sg_photo} with $t_{\rm v,in} = 0.01 \, \text{Myr}$ and $\tw = 5 \, \text{Myr}$.  The left panels of Fig.~\ref{fig:photo_damp} show the evolution of various angles and frequencies over the disk's lifetime, while the right panels zoom in around $t \approx \tw$.  The top right panel shows $\bs$ and $\ld$ decouple first, since $\theta_{\rm sd} > \theta_{\rm pd}, \theta_{\rm sp}$ after $\tw$.  The bottom panel of Fig.~\ref{fig:photo_damp} shows that because $\bompd \gtrsim \omps$ directly after $\tw$, $\blp$ remains strongly coupled to $\ld$, while $\bomsd$ has fallen well below $\omsp$ and $\bomdb$ in magnitude.  Because $\omsp \gtrsim \bomdb$, $\bs$ decouples from $\ld$ and begins to adiabatically follow $\blp$.  The adiabatic trailing of $\bs$ around $\blp$ directly following $\tw$ suppresses any $\theta_{\rm sp}$ excitation (top right panel of Fig.~\ref{fig:photo_damp}).  The top left panel shows by the end of the disk's lifetime, $\tg_{\rm sp}$ is not excited, the expected outcome since $\omsp \gtrsim \bomdb$ (bottom left panel of Fig.~\ref{fig:photo_damp}).  The star-planet system ends up decoupled from the disk, which continues to precess around the binary's orbital angular momentum vector $\blb$.

Figure~\ref{fig:photo_excite} is identical to Fig.~\ref{fig:photo_damp}, except $\ap = 8 \, \rin = 0.294 \, \text{au}$ so the system falls into the $\omsp \lesssim \bomdb$ regime.  The bottom right panel of Fig.~\ref{fig:photo_excite} shows for a brief amount of time following $\tw$, $\blp$ is still strongly coupled to $\ld$ but not to $\bs$ ($\bompd \gtrsim \omps$), while $\bs$ quickly decouples from $\ld$ and couples to $\blp$ ($\omsp \gtrsim \bomsd$).  Because $\omsp \lesssim \bomdb$, $\bs$ cannot adiabatically track $\blp$ due to its rapid precession around $\blb$ through the strong coupling of $\blp$ to $\ld$, explaining why $\theta_{\rm sd} > \theta_{\rm pd}, \theta_{\rm sp}$ when $t > \tw$ (top right panel of Fig.~\ref{fig:photo_excite}).  Some time after $\tw$, $\theta_{\rm sp}$ is excited to large values (top left panel of Fig.~\ref{fig:photo_excite}).  After $t \gtrsim \tw + \text{few} \times \tv$, $\bompd$ falls well below $\omps$ (bottom left panel of Fig.~\ref{fig:photo_excite}), causing $\blp$ to couple strongly with $\bs$, explaining why $\theta_{\rm sp} \sim \text{constant}$ soon after its excitation at $t \approx \tw$.

These two examples show that although the planet-star-disk-binary system's dynamics following $\tw$ is complex, the key criterion for exciting stellar obliquities is $\omsp \lesssim \bomdb$ at $t \approx \tw$.  This is similar to the criterion discussed in Section~\ref{sec:EarlyInSitu} [see Eqs.~\eqref{eq:ap_lower_gen}-\eqref{eq:ap_lower_3/2}].

\section{Discussion}
\label{sec:Discuss}

\subsection{Observational Implications}
\label{sec:ObsImps}

Due to the prevalence of circumstellar disks in binary systems misaligned with the binary's orbital plane (e.g. \citealt{Stapelfeldt(1998),JensenAkeson(2014)}), star-disk-binary interactions have been suggested to generate primordial spin-orbit misalignments in exoplanetary systems \citep{Batygin(2012),BatyginAdams(2013),Lai(2014),SpaldingBatygin(2014)}.  For a wide range of disk/binary parameters, the star-disk-binary system naturally passes through a secular resonance  during the disk's evolution, in which the precession rate of the stellar spin driven by the disk ($\bomsd$) matches the precession rate of the disk driven by the binary companion ($\bomdb$).  When the system passes through this secular resonance, a significant misalignment between the stellar spin and disk axis is generated, even for systems with low disk-binary inclinations (see Fig.~\ref{fig:tg_stand}).  Because this mechanism is so robust, the effects of accretion and magnetic interactions have been invoked to damp the spin-disk misalignment \citep{Lai(2014),SpaldingBatygin(2015)}, and to explain the observed correlation between stellar effective temperatures and obliquities \citep{Winn(2010),Albrecht(2012),Mazeh(2015),LiWinn(2016),Winn(2017)}.

We have shown that when a giant planet forms or migrates to a semi-major axis $\ap$ where  the precession rate of the spinning star around the planet exceeds the precession rate of the disk around the binary companion [$\omsp \gtrsim \bomdb$, see Eqs.~\eqref{eq:omsp} and~\eqref{eq:bomdb}], stellar obliquity excitation may be reduced or completely suppressed.  The excitation is reduced when the planet-star-disk-binary system undergoes secular resonance before the planet migrates near the inner edge of the protoplanetary disk (Fig.~\ref{fig:Mig2}).  The obliquity excitation is completely suppressed when the planet forms or migrates to a semi-major axis too close to its host star before the system has a chance to experience secular resonance (Figs.~\ref{fig:EarlyInSitu} and~\ref{fig:Mig1}).   The obliquity excitation may be maintained if the planet accretes a significant amount of mass non-locally over sufficiently short timescales (Fig.~\ref{fig:LateInSitu}).  We find in order for star-disk-binary interactions to generate significant misalignment between a planet and its host star ($\tg_{\rm sp} \gtrsim 30^\circ$), the planet must lie on an orbit where the disk-binary precession frequency exceeds the star-planet precession frequency ($\bomdb \gtrsim \omsp$), or the system's second secular resonance (when $\omsp \sim \bomdb$) must be crossed quickly.  Rapidly clearing the protoplanetary disk's inner region via photoionization does not modify this criterion (Sec.~\ref{sec:PlanetPhoto}). Assuming the gas-giant's growth is sufficiently slow, a lower bound may be placed on the semi-major axis $\ap$ of a planet [see Eqs.~\eqref{eq:ap_lower_gen}-\eqref{eq:ap_lower_3/2}] which may have primordial spin-orbit misalignment generated via star-disk-binary interactions (assuming $p = 1$):
\be
\ap \gtrsim 0.24 \left( \frac{2 \kq}{\ks} \right)^{1/3} \left( \frac{\Mp}{1 \, \MJ} \right)^{1/3} \left( \frac{\bOmst}{0.1} \right)^{1/3} \frac{\bRs^{1/2} \bab}{\bMb^{1/3} \brout^{1/2}} \, \text{au}.
\label{eq:ap_lower_1}
\ee
 Eq.~\eqref{eq:ap_lower_1} depends very weakly on the surface density power-law index $p$ [Eq.~\eqref{eq:ap_lower_gen}].

\subsubsection{Misalignments of Hot Jupiter Systems}
\label{sec:MisalignedHJ}

Equation~\eqref{eq:ap_lower_1} [see also Eqs.~\eqref{eq:ap_lower_gen}-\eqref{eq:ap_lower_3/2}] shows that while ``cold'' Jupiters (planets with mass $\Mp \sim 1 \, \MJ$ and semi-major axis $\ap \gtrsim 1 \, \text{au}$) and close-in earth-mass planets can experience primordial misalignment excitations from binary companions, hot Jupiters (HJ) may have this primordial misalignment reduced or completely suppressed.  Most systems with significant stellar obliquities detected via the Rossiter-McLaughlin effect are HJs with semi-major axis $\ap \lesssim 0.1 \, \text{au}$ \citep{WinnFabrycky(2015),Triaud(2017)}.  Thus our work shows that HJs with star-planet inclinations $\tg_{\rm sp} \gtrsim 30^\circ$ are unlikely to have developed these misalignments through planet-star-disk-binary interactions.  

Other ways to primordially misalign the stellar spin axis with the planet's orbital angular momentum include molecular cloud turbulence during the formation of circumstellar disks \citep{Bate(2000),Bate(2010),Fielding(2015)} and magnetic star-disk interactions \citep{Lai(2011),FoucartLai(2011)}.  For molecular cloud turbulence to generate spin-orbit misalignment primordially, the in-falling cloud material must cause the disk's orbital angular momentum vector to vary on timescales less than the precession period of the stellar spin around the disk \citep{Spalding(2014)}.  Our work shows that the location and time of massive planet formation is also relevant for how efficiently molecular cloud turbulence may generate primordial spin-orbit misalignments.

\subsubsection{Kepler Multi-planet Systems}
\label{sec:KeplerMultis}

Observational evidence suggesting multi-planet systems discovered by \textit{Kepler} have low stellar obliquities is beginning to mount \citep{Albrecht(2013),Winn(2017)}.  Equation~\eqref{eq:ap_lower_1} shows that such systems may experience primordial obliquity excitation by inclined binary companions, depending on the mass of the planet and the semi-major axis of the companion.  If HJs form in multi-planetary systems, but are subsequently disrupted or engulfed by the host star (e.g. by stellar tides), the hot Jupiter's presence before the protoplanetary disk dissipates may ``protect" multi-planet systems from primordial spin-orbit misalignment excitation in the presence of an inclined binary companion. The occurrence rate of HJs is $\lesssim 1 \%$ (e.g. \citealt{Marcy(2005),Gould(2006),Cumming(2008),Howard(2010),Howard(2012),BaylissSackett(2011),Wright(2012)}), the occurrence rates of Jovian-mass planets orbiting interior to $\sim \text{few} \, \text{au}$ is $\sim 5-10 \%$ (e.g. \citealt{Cumming(2008),ClantonGaudi(2016)}), and potentially as high as $\sim 50 \%$ for long-period (semi-major axis $\sim \text{few} - 100 \, \text{au}$) giant planets (e.g. \citealt{ClantonGaudi(2016),Foreman-Mackey(2016),Vigan(2017)}).  If the population of ``destroyed'' HJs is comparable to the fraction of long-period massive planets, a significant fraction of low-obliquity ($\tg_{\rm sp} \approx 0^\circ$) exoplanetary systems may be explained through this mechanism.  The multi-planet system WASP-47, containing a HJ with two low-mass planet ``neighbors'' \citep{Becker(2015)} and a stellar obliquity $\tg_{\rm sp} = 0 \pm 24^\circ$ consistent with 0 \citep{Sanchis-Ojeda(2015)}, may be an example of a multi-planet system with a HJ ``protector'' which has survived to the present day.

\subsection{Theoretical Uncertainties}
\label{sec:TheoryUncertain}

In Section~\ref{sec:Photo}, we considered a simple model for non-homologous surface density evolution, parameterizing the UV-switch model of photoevaporation \citep{Clarke(2001)}.  Our model assumes the critical radius $\rc$ (where the disk's photo-ionization rate is comparable to the disk's viscous depletion rate) is fixed in time.  In reality, photoevaporation forces $\rc$ to expand in time shortly ($\lesssim 10^5 \, \text{years}$) after the inner disk is viscously depleted onto the central star, and expands outward over a timescale of $\sim 10^5 \, \text{years}$ \citep{Alexander(2006),Owen(2010)}.  There is not a simple way to relate this timescale to $\tw$ (when the inner disk begins to clear) and $t_{\rm v,in}$ (the timescale over which the inner disk is depleted).  In addition, the expansion of $\rc$ may accelerate due to a dynamical instability termed ``thermal sweeping" \citep{Owen(2012),Haworth(2016)}.  Including the expansion of $\rc$ will work to reduce the magnitudes of the precession rate of the stellar spin around the outer disk [$\bomsdg$, Eq.~\eqref{eq:bomsdg}] and the precession rate of the planet about the outer disk [$\bompdg$, Eq.~\eqref{eq:bompdg}], which already become insignificant shortly after the inner disk is viscously drained ($t \gtrsim \tw + \text{few} \times \tv$, see Figs.~\ref{fig:photo_damp}-\ref{fig:photo_excite}).  Since the excitation of stellar obliquities with a planet inside the disk's inner cavity depends mainly on the planet-star-disk-binary properties when the inner disk is drained ($t \approx \tw$), inclusion of an expanding $\rc$ will not change the main results of this paper.

Other models exist which cause the disk surface density to evolve non-homologously.  For instance, \cite{RussoThompson(2015a),RussoThompson(2015b)} considered the evolution of protoplanetary disks through magnetorotational instability \citep{BalbusHawley(1991)} driven turbulence, seeded by magnetized stellar winds.  Because the magnetic field from the star is larger near the inner truncation radius, the inner region of the disk is more turbulent than the outer region.  This model for turbulence in disks around T-Tauri stars results in the inner ($r\lesssim \text{few} \ \text{au}$) disk accreting in less than a few Myr.  Investigating the effect of such non-homologous disk evolution models is outside the scope of this paper.

In this work, we have assumed that the disk is flat (i.e. $\ld$ is independent of radius).  Disk warping may change the mutual inclinations of planets formed in star-disk-binary systems, and align the stellar spin, disk, and binary orbital angular momentum axis over viscous timescales.  These issues are addressed in a companion work \citep{ZanazziLai(2017b)}.  Disk warping may also affect the planet-disk interaction of massive planets, explored in \cite{LubowOgilvie(2001)}.

\section{Conclusions}
\label{sec:Conc}

In this work, we have studied how the formation of a massive planet orbiting close to its host star affects the generation of primordial misalignment between the stellar spin and the orbital angular momentum axes of the planet and disk in the presence of an inclined external binary companion.  We find that when a protoplanetary disk's inner cavity is rapidly cleared by photoevaporation before secular resonance occurs, the star-disk misalignment is reduced (Sec.~\ref{sec:Photo}, Fig.~\ref{fig:tg_photo}).  More importantly, when a giant planet (hot Jupiter) forms or migrates early to the near vicinity of the host star, it becomes strongly coupled to the star, preventing any significant excitation of spin-orbit misalignment.  Specifically, we find (Sec.~\ref{sec:DynPSDB}):
\begin{enumerate}
\item If a hot Jupiter forms early in-situ, spin-orbit misalignment is \textit{completely suppressed} since the star-planet precession frequency $\omsp$ [Eq.~\eqref{eq:omsp}] always exceeds the disk-binary precession frequency $\bomdb$ [Eq.~\eqref{eq:bomdb}].
\item If a hot Jupiter forms late in-situ and accretes most of its mass from viscously advected disk gas, spin-orbit misalignment is \textit{significantly reduced} when the planet's mass grows over a timescale sufficiently longer than the disk-binary precession period (Fig.~\ref{fig:LateInSitu}).  Spin-orbit misalignment can be maintained if the planet's mass growth timescale [Eq.~\eqref{eq:t_grow}] is shorter than the disk-binary precession period (Fig.~\ref{fig:LateInSitu}).
\item If a giant planet forms in the outer region of the protoplanetary disk, and migrates in via Type-II migration, the excitation of spin-orbit misalignments depends on the migration history of the planet relative to the time of the secular resonance.  If the planet migrates to a semi-major axis $\ap$ such that the star-planet precession frequency exceeds the disk-binary precession frequency ($\omsp \gtrsim \bomdb$) \textit{before} secular resonance, spin-orbit misalignment is \textit{completely suppressed} (Fig.~\ref{fig:Mig1}, top left panel).  If the planet migration occurs after secular resonance, the stellar obliquity is significantly reduced when the planet migrates to a close-in orbit which satisfies $\omsp \gtrsim \bomdb$ (Fig.~\ref{fig:Mig2}).
\item If the giant planet is left in the photoevaporated inner disk cavity before secular resonance occurs in the star-disk-binary system, spin-orbit misalignment is \textit{completely suppressed}, unless the star-planet precession frequency is less than the disk-binary precession frequency ($\omsp \lesssim \bomdb$) when the inner disk is cleared (Figs.~\ref{fig:photo_damp}-\ref{fig:photo_excite}).
\end{enumerate}

Overall,  our work shows that regardless of the complication of disk evolution, significant stellar obliquities can be generated only when the planet forms at an orbital separation where the star-planet precession frequency exceeds the disk-binary precession frequency ($\omsp \lesssim \bomdb$), or the planet forms quickly after the system undergoes secular resonance.  This places a lower bound on a slowly forming planet's semi-major axis such that primordial spin-orbit misalignment can be generated via star-disk-binary interactions [Eqs.~\eqref{eq:ap_lower_gen},\eqref{eq:ap_lower_3/2} or~\eqref{eq:ap_lower_1}].  Hot Jupiters do not satisfy this bound, and thus may not acquire significant spin-orbit misalignment through this mechanism, depending on how the planet accreted its mass (Sec.~\ref{sec:MisalignedHJ}).  On the other hand, ``cold Jupiters'' and close-in earth-mass planets may experience excitation of spin-orbit misalignments under appropriate conditions.  If hot Jupiters form in multi-planet systems, they may protect the system from suffering primordial spin-orbit misalignments in the presence of an inclined binary companion (Sec.~\ref{sec:KeplerMultis}).

\section*{Acknowledgements}

JZ thanks Doug Lin, Christopher Spalding, Konstantin Batygin, and Eve Lee for helpful discussions.  This work has been supported in part by NASA grants
NNX14AG94G and NNX14AP31G, and NSF grant AST-
1715246.  JZ is supported by a NASA Earth and Space Sciences Fellowship in Astrophysics.


\begin{thebibliography}{99}

\bibitem[\protect\citeauthoryear{Albrecht et al.}{2013}]{Albrecht(2013)} Albrecht S., Winn J.~N., Marcy G.~W., Howard A.~W., Isaacson H., Johnson J.~A., 2013, ApJ, 771, 11 

\bibitem[Albrecht et al.(2012)]{Albrecht(2012)} Albrecht, S., Winn, J.~N., Johnson, J.~A., et al.\ 2012, ApJ, 757, 18

\bibitem[\protect\citeauthoryear{Alexander et al.}{2014}]{Alexander(2014)} Alexander R., Pascucci I., Andrews S., Armitage P., Cieza L., 2014, prpl.conf, 475

\bibitem[\protect\citeauthoryear{Alexander, Clarke  \& Pringle}{2006}]{Alexander(2006)} Alexander R.~D., Clarke C.~J., Pringle J.~E., 2006, MNRAS, 369, 229 

\bibitem[\protect\citeauthoryear{Anderson, Storch \& Lai}{2016}]{Anderson(2016)} Anderson K.~R., Storch N.~I., Lai D., 2016, MNRAS, 456, 3671 

\bibitem[\protect\citeauthoryear{Ayliffe \& Bate}{2009}]{AyliffeBate(2009)} Ayliffe B.~A., Bate M.~R., 2009, MNRAS, 393, 49 

\bibitem[Balbus \& Hawley(1991)]{BalbusHawley(1991)} Balbus, S.~A., \& Hawley, J.~F.\ 1991, ApJ, 376, 214 

\bibitem[\protect\citeauthoryear{Bate, Lodato \& Pringle}{2010}]{Bate(2010)} Bate M.~R., Lodato G., Pringle J.~E., 2010, MNRAS, 401, 1505

\bibitem[\protect\citeauthoryear{Bate et al.}{2000}]{Bate(2000)} Bate M.~R., Bonnell I.~A., Clarke C.~J., Lubow S.~H., Ogilvie G.~I., Pringle J.~E., Tout C.~A., 2000, MNRAS, 317, 773 

\bibitem[\protect\citeauthoryear{Bayliss \& Sackett}{2011}]{BaylissSackett(2011)} Bayliss D.~D.~R., Sackett P.~D., 2011, ApJ, 743, 103 

\bibitem[\protect\citeauthoryear{Batygin}{2012}]{Batygin(2012)} Batygin K., 2012, Natur, 491, 418 

\bibitem[\protect\citeauthoryear{Batygin \& Adams}{2013}]{BatyginAdams(2013)} Batygin K., Adams F.~C., 2013, ApJ, 778, 169 

\bibitem[\protect\citeauthoryear{Batygin, Bodenheimer \& Laughlin}{2016}]{Batygin(2016)} Batygin K., Bodenheimer P.~H., Laughlin G.~P., 2016, ApJ, 829, 114 

\bibitem[\protect\citeauthoryear{Beaug{\'e} \& Nesvorn{\'y}}{2012}]{BeaugeNesvorny(2012)} Beaug{\'e} C., Nesvorn{\'y} D., 2012, ApJ, 751, 119 

\bibitem[\protect\citeauthoryear{Becker et al.}{2015}]{Becker(2015)} Becker J.~C., Vanderburg A., Adams F.~C., Rappaport S.~A., Schwengeler H.~M., 2015, ApJ, 812, L18 

\bibitem[\protect\citeauthoryear{Ben{\'{\i}}tez-Llambay et al.}{2015}]{Benitez-Llambay(2015)} Ben{\'{\i}}tez-Llambay P., Masset F., Koenigsberger G., Szul{\'a}gyi J., 2015, Natur, 520, 63

\bibitem[\protect\citeauthoryear{Bitsch et al.}{2013}]{Bitsch(2013)} Bitsch B., Crida A., Libert A.-S., Lega E., 2013, A\&A, 555, A124 

\bibitem[\protect\citeauthoryear{Boley, Granados Contreras \& Gladman}{2016}]{Boley(2016)} Boley A.~C., Granados Contreras A.~P., Gladman B., 2016, ApJ, 817, L17 

\bibitem[\protect\citeauthoryear{Bryan et al.}{2016}]{2016ApJ...827..100B} Bryan M.~L., Bowler B.~P., Knutson H.~A., Kraus A.~L., Hinkley S., Mawet D., Nielsen E.~L., Blunt S.~C., 2016, ApJ, 827, 100 

\bibitem[\protect\citeauthoryear{Chametla et al.}{2017}]{Chametla(2017)} Chametla R.~O., S{\'a}nchez-Salcedo F.~J., Masset F.~S., Hidalgo-G{\'a}mez A.~M., 2017, MNRAS, 468, 4610 

\bibitem[\protect\citeauthoryear{Chiang \& Laughlin}{2013}]{ChaingLaughlin(2013)} Chiang E., Laughlin G., 2013, MNRAS, 431, 3444 

\bibitem[\protect\citeauthoryear{Clanton \& Gaudi}{2016}]{ClantonGaudi(2016)} Clanton C., Gaudi B.~S., 2016, ApJ, 819, 125 

\bibitem[Clarke et al.(2001)]{Clarke(2001)} Clarke, C.~J., Gendrin, A., \& Sotomayor, M.\ 2001, MNRAS, 328, 485

\bibitem[\protect\citeauthoryear{Cresswell et al.}{2007}]{Cresswell(2007)} Cresswell P., Dirksen G., Kley W., Nelson R.~P., 2007, A\&A, 473, 329 

\bibitem[\protect\citeauthoryear{Cumming et al.}{2008}]{Cumming(2008)} Cumming A., Butler R.~P., Marcy G.~W., Vogt S.~S., Wright J.~T., Fischer D.~A., 2008, PASP, 120, 531 

\bibitem[\protect\citeauthoryear{D'Angelo \& Lubow}{2008}]{D'AngeloLubow(2008)} D'Angelo G., Lubow S.~H., 2008, ApJ, 685, 560-583 

\bibitem[\protect\citeauthoryear{David et al.}{2016}]{David(2016)} David T.~J., et al., 2016, Natur, 534, 658 

\bibitem[\protect\citeauthoryear{Dawson, Murray-Clay \& Johnson}{2015}]{Dawson(2015)} Dawson R.~I., Murray-Clay R.~A., Johnson J.~A., 2015, ApJ, 798, 66

\bibitem[\protect\citeauthoryear{Donati et al.}{2016}]{Donati(2016)} Donati J.~F., et al., 2016, Natur, 534, 662 

\bibitem[\protect\citeauthoryear{Duffell et al.}{2014}]{Duffell(2014)} Duffell P.~C., Haiman Z., MacFadyen A.~I., D'Orazio D.~J., Farris B.~D., 2014, ApJ, 792, L10 

\bibitem[\protect\citeauthoryear{D{\"u}rmann \& Kley}{2015}]{DurmannKley(2015)} D{\"u}rmann C., Kley W., 2015, A\&A, 574, A52 

\bibitem[\protect\citeauthoryear{Fabrycky \& Tremaine}{2007}]{FabryckyTremaine(2007)} Fabrycky D., Tremaine S., 2007, ApJ, 669, 1298 

\bibitem[\protect\citeauthoryear{Fielding et al.}{2015}]{Fielding(2015)} Fielding D.~B., McKee C.~F., Socrates A., Cunningham A.~J., Klein R.~I., 2015, MNRAS, 450, 3306

\bibitem[\protect\citeauthoryear{Foreman-Mackey et al.}{2016}]{Foreman-Mackey(2016)} Foreman-Mackey D., Morton T.~D., Hogg D.~W., Agol E., Sch{\"o}lkopf B., 2016, AJ, 152, 206 

\bibitem[\protect\citeauthoryear{Foucart \& Lai}{2011}]{FoucartLai(2011)} Foucart F., Lai D., 2011, MNRAS, 412, 2799 

\bibitem[\protect\citeauthoryear{Goldreich \& Tremaine}{1979}]{GoldreichTremaine(1979)} Goldreich P., Tremaine S., 1979, ApJ, 233, 857 

\bibitem[\protect\citeauthoryear{Gould et al.}{2006}]{Gould(2006)} Gould A., Dorsher S., Gaudi B.~S., Udalski A., 2006, AcA, 56, 1 

\bibitem[\protect\citeauthoryear{Hahn}{2003}]{Hahn(2003)} Hahn J.~M., 2003, ApJ, 595, 531 

\bibitem[\protect\citeauthoryear{Hamers \& Portegies Zwart}{2016}]{Hamers(2016)} Hamers A.~S., Portegies Zwart S.~F., 2016, MNRAS, 459, 2827 

\bibitem[\protect\citeauthoryear{Haworth, Clarke \& Owen}{2016}]{Haworth(2016)} Haworth T.~J., Clarke C.~J., Owen J.~E., 2016, MNRAS, 457, 1905 

\bibitem[\protect\citeauthoryear{H{\'e}brard et al.}{2008}]{Hebrard(2008)} H{\'e}brard G., et al., 2008, A\&A, 488, 763 

\bibitem[Hollenbach et al.(1994)]{Hollenbach(1994)} Hollenbach, D., Johnstone, D., Lizano, S., \& Shu, F.\ 1994, ApJ, 428, 654 

\bibitem[\protect\citeauthoryear{Howard et al.}{2012}]{Howard(2012)} Howard A.~W., et al., 2012, ApJS, 201, 15 

\bibitem[\protect\citeauthoryear{Howard et al.}{2010}]{Howard(2010)} Howard A.~W., et al., 2010, Sci, 330, 653 

\bibitem[\protect\citeauthoryear{Ida \& Lin}{2004}]{IdaLin(2004)} Ida S., Lin D.~N.~C., 2004, ApJ, 604, 388 

\bibitem[\protect\citeauthoryear{Jensen \& Akeson}{2014}]{JensenAkeson(2014)} Jensen E.~L.~N., Akeson R., 2014, Natur, 511, 567 

\bibitem[\protect\citeauthoryear{Jim{\'e}nez \& Masset}{2017}]{JimenezMasset(2017)} Jim{\'e}nez M.~A., Masset F.~S., 2017, MNRAS, 471, 4917 

\bibitem[\protect\citeauthoryear{Kley \& Nelson}{2012}]{KleyNelson(2012)} Kley W., Nelson R.~P., 2012, ARA\&A, 50, 211 

\bibitem[Lai(2014)]{Lai(2014)} Lai, D.\ 2014, MNRAS, 440, 3532 

\bibitem[Lai et al.(2011)]{Lai(2011)} Lai, D., Foucart, F., \& Lin, D.~N.~C.\ 2011, MNRAS, 412, 2790 

\bibitem[\protect\citeauthoryear{Lai, Rasio \& Shapiro}{1993}]{Lai(1993)} Lai D., Rasio F.~A., Shapiro S.~L., 1993, ApJS, 88, 205

\bibitem[\protect\citeauthoryear{Li \& Winn}{2016}]{LiWinn(2016)} Li G., Winn J.~N., 2016, ApJ, 818, 5 

\bibitem[\protect\citeauthoryear{Lin, Bodenheimer \& Richardson}{1996}]{Lin(1996)} Lin D.~N.~C., Bodenheimer P., Richardson D.~C., 1996, Natur, 380, 606 

\bibitem[\protect\citeauthoryear{Lin \& Papaloizou}{1993}]{LinPapaloizou(1993)} Lin D.~N.~C., Papaloizou J.~C.~B., 1993, prpl.conf, 749 

\bibitem[\protect\citeauthoryear{Lin \& Papaloizou}{1985}]{LinPapaloizou(1985)} Lin D.~N.~C., Papaloizou J., 1985, prpl.conf, 981 

\bibitem[\protect\citeauthoryear{Lubow \& D'Angelo}{2006}]{LubowD'Angelo(2006)} Lubow S.~H., D'Angelo G., 2006, ApJ, 641, 526 

\bibitem[\protect\citeauthoryear{Lubow \& Martin}{2016}]{LubowMartin(2016)} Lubow S.~H., Martin R.~G., 2016, ApJ, 817, 30 

\bibitem[\protect\citeauthoryear{Lubow \& Ogilvie}{2001}]{LubowOgilvie(2001)} Lubow S.~H., Ogilvie G.~I., 2001, ApJ, 560, 997 

\bibitem[\protect\citeauthoryear{Lubow \& Ogilvie}{2000}]{LubowOgilvie(2000)} Lubow S.~H., Ogilvie G.~I., 2000, ApJ, 538, 326 

\bibitem[\protect\citeauthoryear{Marcy et al.}{2005}]{Marcy(2005)} Marcy G., Butler R.~P., Fischer D., Vogt S., Wright J.~T., Tinney C.~G., Jones H.~R.~A., 2005, PThPS, 158, 24 

\bibitem[\protect\citeauthoryear{Martin et al.}{2016}]{Martin(2016)} Martin R.~G., Lubow S.~H., Nixon C., Armitage P.~J., 2016, MNRAS, 458, 4345 

\bibitem[\protect\citeauthoryear{Masset}{2017}]{Masset(2017)} Masset F.~S., 2017, arXiv, arXiv:1708.09807 

\bibitem[\protect\citeauthoryear{Masset \& Velasco Romero}{2017}]{MassetVelascoRomero(2017)} Masset F.~S., Velasco Romero D.~A., 2017, MNRAS, 465, 3175 

\bibitem[\protect\citeauthoryear{Matsakos \& K{\"o}nigl}{2017}]{MatsakosKonigl(2017)} Matsakos T., K{\"o}nigl A., 2017, AJ, 153, 60 

\bibitem[\protect\citeauthoryear{Mazeh et al.}{2015}]{Mazeh(2015)} Mazeh T., Perets H.~B., McQuillan A., Goldstein E.~S., 2015, ApJ, 801, 3 

\bibitem[\protect\citeauthoryear{McNally et al.}{2017}]{McNally(2017)} McNally C.~P., Nelson R.~P., Paardekooper S.-J., Gressel O., Lyra W., 2017, MNRAS, 472, 1565 

\bibitem[\protect\citeauthoryear{Mu{\~n}oz, Lai \& Liu}{2016}]{Munoz(2016)} Mu{\~n}oz D.~J., Lai D., Liu B., 2016, MNRAS, 460, 1086 

\bibitem[\protect\citeauthoryear{Nagasawa, Ida, \& Bessho}{2008}]{Nagasawa(2008)} Nagasawa M., Ida S., Bessho T., 2008, ApJ, 678, 498-508 

\bibitem[\protect\citeauthoryear{Naoz, Farr \& Rasio}{2012}]{Naoz(2012)} Naoz S., Farr W.~M., Rasio F.~A., 2012, ApJ, 754, L36 

\bibitem[\protect\citeauthoryear{Narita et al.}{2009}]{Narita(2009)} Narita N., Sato B., Hirano T., Tamura M., 2009, PASJ, 61, L35 

\bibitem[Owen(2016)]{Owen(2016)} Owen, J.~E.\ 2016, PASA, 33, e005 

\bibitem[\protect\citeauthoryear{Owen, Clarke \& Ercolano}{2012}]{Owen(2012)} Owen J.~E., Clarke C.~J., Ercolano B., 2012, MNRAS, 422, 1880 

\bibitem[\protect\citeauthoryear{Owen et al.}{2010}]{Owen(2010)} Owen J.~E., Ercolano B., Clarke C.~J., Alexander R.~D., 2010, MNRAS, 401, 1415 

\bibitem[\protect\citeauthoryear{Papaloizou \& Nelson}{2005}]{PapaloizouNelson(2005)} Papaloizou J.~C.~B., Nelson R.~P., 2005, A\&A, 433, 247 

\bibitem[\protect\citeauthoryear{Papaloizou \& Lin}{1995}]{PapaloizouLin(1995)} Papaloizou J.~C.~B., Lin D.~N.~C., 1995, ApJ, 438, 841 

\bibitem[\protect\citeauthoryear{Petrovich}{2015}]{Petrovich(2015)} Petrovich C., 2015, ApJ, 805, 75 

\bibitem[\protect\citeauthoryear{Pollack et al.}{1996}]{Pollack(1996)} Pollack J.~B., Hubickyj O., Bodenheimer P., Lissauer J.~J., Podolak M., Greenzweig Y., 1996, Icar, 124, 62 

\bibitem[\protect\citeauthoryear{Rafikov}{2017}]{Rafikov(2017)} Rafikov R.~R., 2017, ApJ, 837, 163 

\bibitem[\protect\citeauthoryear{Rogers, Lin, \& Lau}{2012}]{Rogers(2012)} Rogers T.~M., Lin D.~N.~C., Lau H.~H.~B., 2012, ApJ, 758, L6

\bibitem[Russo \& Thompson(2015a)]{RussoThompson(2015a)} Russo, M., \& Thompson, C.\ 2015, ApJ, 813, 81 

\bibitem[Russo \& Thompson(2015b)]{RussoThompson(2015b)} Russo, M., \& Thompson, C.\ 2015, ApJ, 815, 38 

\bibitem[\protect\citeauthoryear{Sanchis-Ojeda et al.}{2015}]{Sanchis-Ojeda(2015)} Sanchis-Ojeda R., et al., 2015, ApJ, 812, L11 

\bibitem[\protect\citeauthoryear{Schlaufman \& Winn}{2016}]{SchlaufmanWinn(2016)} Schlaufman K.~C., Winn J.~N., 2016, ApJ, 825, 62 

\bibitem[\protect\citeauthoryear{Spalding \& Batygin}{2017}]{SpaldingBatygin(2017)} Spalding C., Batygin K., 2017, AJ, 154, 93 

\bibitem[Spalding \& Batygin(2015)]{SpaldingBatygin(2015)} Spalding, C., \& Batygin, K.\ 2015, ApJ, 811, 82

\bibitem[\protect\citeauthoryear{Spalding \& Batygin}{2014}]{SpaldingBatygin(2014)} Spalding C., Batygin K., 2014, ApJ, 790, 42 


\bibitem[\protect\citeauthoryear{Spalding, Batygin \& Adams}{2014}]{Spalding(2014)} Spalding C., Batygin K., Adams F.~C., 2014, ApJ, 797, L29 

\bibitem[\protect\citeauthoryear{Stapelfeldt et al.}{1998}]{Stapelfeldt(1998)} Stapelfeldt K.~R., Krist J.~E., M{\'e}nard F., Bouvier J., Padgett D.~L., Burrows C.~J., 1998, ApJ, 502, L65 

\bibitem[\protect\citeauthoryear{Storch, Lai \& Anderson}{2017}]{Storch(2017)} Storch N.~I., Lai D., Anderson K.~R., 2017, MNRAS, 465, 3927 

\bibitem[\protect\citeauthoryear{Storch \& Lai}{2015}]{StorchLai(2015)} Storch N.~I., Lai D., 2015, MNRAS, 448, 1821 

\bibitem[\protect\citeauthoryear{Storch, Anderson \& Lai}{2014}]{Storch(2014)} Storch N.~I., Anderson K.~R., Lai D., 2014, Sci, 345, 1317 

\bibitem[\protect\citeauthoryear{Tanaka \& Ward}{2004}]{TanakaWard(2004)} Tanaka H., Ward W.~R., 2004, ApJ, 602, 388 

\bibitem[\protect\citeauthoryear{Tanaka, Takeuchi, \& Ward}{2002}]{Tanaka(2002)} Tanaka H., Takeuchi T., Ward W.~R., 2002, ApJ, 565, 1257 

\bibitem[\protect\citeauthoryear{Tanigawa \& Tanaka}{2016}]{TanigawaTanaka(2016)} Tanigawa T., Tanaka H., 2016, ApJ, 823, 48 

\bibitem[\protect\citeauthoryear{Thies et al.}{2011}]{Thies(2011)} Thies I., Kroupa P., Goodwin S.~P., Stamatellos D., Whitworth A.~P., 2011, MNRAS, 417, 1817 

\bibitem[\protect\citeauthoryear{Triaud}{2017}]{Triaud(2017)} Triaud A.~H.~M.~J., 2017, arXiv, arXiv:1709.06376 

\bibitem[\protect\citeauthoryear{Triaud et al.}{2010}]{Triaud(2010)} Triaud A.~H.~M.~J., et al., 2010, A\&A, 524, A25 

\bibitem[\protect\citeauthoryear{Vigan et al.}{2017}]{Vigan(2017)} Vigan A., et al., 2017, A\&A, 603, A3 

\bibitem[\protect\citeauthoryear{Ward}{1981}]{Ward(1981)} Ward W.~R., 1981, Icar, 47, 234 

\bibitem[\protect\citeauthoryear{Weidenschilling}{1977}]{Weidenschilling(1977)} Weidenschilling S.~J., 1977, Ap\&SS, 51, 153 

\bibitem[\protect\citeauthoryear{Williams \& Cieza}{2011}]{WilliamsCieza(2011)} Williams J.~P., Cieza L.~A., 2011, ARA\&A, 49, 67 

\bibitem[\protect\citeauthoryear{Winn et al.}{2017}]{Winn(2017)} Winn J.~N., et al., 2017, arXiv, arXiv:1710.04530 

\bibitem[\protect\citeauthoryear{Winn \& Fabrycky}{2015}]{WinnFabrycky(2015)} Winn J.~N., Fabrycky D.~C., 2015, ARA\&A, 53, 409 

\bibitem[\protect\citeauthoryear{Winn et al.}{2010}]{Winn(2010)} Winn J.~N., Fabrycky D., Albrecht S., Johnson J.~A., 2010, ApJ, 718, L145 

\bibitem[\protect\citeauthoryear{Winn et al.}{2009}]{Winn(2009)} Winn J.~N., Johnson J.~A., Albrecht S., Howard A.~W., Marcy G.~W., Crossfield I.~J., Holman M.~J., 2009, ApJ, 703, L99 

\bibitem[\protect\citeauthoryear{Wright et al.}{2012}]{Wright(2012)} Wright J.~T., Marcy G.~W., Howard A.~W., Johnson J.~A., Morton T.~D., Fischer D.~A., 2012, ApJ, 753, 160 

\bibitem[\protect\citeauthoryear{Wu \& Lithwick}{2011}]{WuLithwick(2011)} Wu Y., Lithwick Y., 2011, ApJ, 735, 109 

\bibitem[\protect\citeauthoryear{Wu \& Murray}{2003}]{WuMurray(2003)} Wu Y., Murray N., 2003, ApJ, 589, 605 

\bibitem[\protect\citeauthoryear{Xiang-Gruess \& Papaloizou}{2013}]{Xiang-GruessPapaloizou(2013)} Xiang-Gruess M., Papaloizou J.~C.~B., 2013, MNRAS, 431, 1320 

\bibitem[\protect\citeauthoryear{Zanazzi \& Lai}{2017a}]{ZanazziLai(2017a)} Zanazzi J.~J., Lai D., 2017a, MNRAS, 464, 3945 

\bibitem[\protect\citeauthoryear{Zanazzi \& Lai}{2017b}]{ZanazziLai(2017b)} Zanazzi J.~J., Lai D., 2017b, arXiv, arXiv:1712.07655 


\end{thebibliography}
\end{document}